\documentclass[useAMS,usenatbib]{mnras}
\usepackage{graphicx}
\usepackage{txfonts}
\usepackage[dvipsnames]{color}
\usepackage{ulem}
\def\aap{A\&A}
\def\apj{ApJ}

\def\apjs{ApJS}

\usepackage{soul}
\usepackage[dvipsnames]{xcolor}

\title[Monte-Carlo simulations of RRMSs]{Monte-Carlo simulations of relativistic radiation mediated shocks: I. photon rich regime}
\author[Ito et al.]{{Hirotaka Ito$^{1,2}$, Amir Levinson$^{3}$, Boris E. Stern$^{4}$ and Shigehiro Nagataki$^{1,2,5}$} \\
$^{1}$Astrophysical Big Bang Laboratory, RIKEN, Saitama 351-0198, Japan; hirotaka.ito@riken.jp\\
$^{2}$Interdisciplinary Theoretical Science (iTHES) Research Group, RIKEN, Wako, Saitama 351-0198, Japan\\
$^{3}$School of Physics \& Astronomy, Tel Aviv University, Tel Aviv 69978,
Israel; Levinson@wise.tau.ac.il\\
$^{4}$Institute for Nuclear Research, Moscow, Russia\\
$^{5}$Interdisciplinary Theoretical \& Mathematical Science Program (iTHEMS), RIKEN, Saitama 351-0198, Japan}

\begin{document}
\date{\today}
\pagerange{000--000} \pubyear{0000}
\maketitle
\label{firstpage}
\begin{abstract}
We explore the physics of relativistic radiation mediated shocks (RRMSs) in the regime where photon advection dominates over photon generation. For this purpose, a novel iterative method for deriving a self-consistent steady-state structure of RRMS is developed, based on a Monte-Carlo code that solves the transfer of photons subject to Compton scattering and pair production/annihilation. Systematic study is performed by imposing various upstream conditions which are characterized by the following three parameters: the photon-to-baryon inertia ratio $\xi_{u *}$, the photon-to-baryon number ratio  $\tilde{n}$, and the shock Lorentz factor $\gamma_u$. We find that the properties of RRMSs vary considerably with these parameters. In particular, while a smooth decline in the velocity, accompanied by a gradual temperature increase is seen for $\xi_{u*} \gg 1$, an efficient bulk Comptonization, that leads to a heating precursor, is found for $\xi_{u*} \lesssim 1$. As a consequence, although particle acceleration is highly inefficient in these shocks, a broad non-thermal spectrum is produced in the latter case. The generation of high energy photons through bulk Comptonization leads, in certain cases, to a copious production of pairs that provide the dominant opacity for Compton scattering. We also find that for certain upstream conditions
 a weak subshock appears within the flow. For a choice of parameters suitable to gamma-ray bursts, the radiation spectrum within the shock is found to be compatible with that of the prompt emission, suggesting that subphotospheric shocks may give rise to the observed non-thermal features despite the absence of accelerated particles.

\end{abstract}


\section{Introduction}
\label{Intro}
Shocks are ubiquitous  in high-energy astrophysics.  They are believed to be the sources of non-thermal photons, 
cosmic-rays and neutrinos observed in extreme astrophysical objects.    Two distinct types of astrophysical shocks
have been identified: "collisionless" shocks, in which dissipation is mediated by collective plasma process, 
and "radiation mediated shocks" (RMS), in which dissipation is governed by Compton scattering and under certain 
conditions also by pair production.  A collisionless shock usually forms in a dilute, optically thin plasma, 
where binary collisions and radiation drag are negligible,  its characteristic width is of the order of the 
plasma skin depth, and it is capable of accelerating 
particles to non-thermal energies in cases where the magnetization of the upstream flow is not too high.
A RMS, on the other hand, forms when a fast shock propagates in an optically thick plasma, its width is of the order of the 
Thomson scattering length, and it cannot accelerate particles to non-thermal energies by virtue of its large width, that 
exceeds any kinetic scale by many orders of magnitudes (see further discussions below regarding this point). 
It is worth noting that while the microphysics of collisionless shocks is poorly understood, and any progress 
 in our understanding of these systems relies heavily on sophisticated plasma (PIC) simulations, the microphysics
of RMS is fully understood, which considerably alleviates the problem.

RMS play a key role in  a variety of astrophysical systems, including shock breakout in supernovae (SNe) and 
low-luminosity GRBs
\citep[e.g.,][for a recent review see Waxman \& Katz 2016]{Colgate1974, KC78, W76, Ch92, RW11, NS10, NS12, SKW11, SN14a, SN14b},
 choked GRB jets 
\citep[][see Nakar 2015 and
Senno et al. 2016 for the implications for neutrino production]{MW01, Na15},
sub-photospheric shocks in GRBs 
\citep{LB08, BML11, LE12, KL14, Ahl15, B17, LBV17},
 and accretion flows into black holes.  
The environments in which the shocks propagate and their velocity vary significantly between the various systems. 
For instance, shocks that are generated by various types of stellar explosions propagate in an unmagnetized, photon poor medium,
and their velocity prior to breakout ranges from sub-relativistic to ultra-relativistic, depending on the type of the progenitor and 
the explosion energy
\citep{NS12}.
Sub-photospheric internal shocks in GRBs, on the other hand,  propagate in a photon rich plasma, conceivably  with a 
non-negligible magnetization, at mildly relativistic speeds.  Consequently, the structure and observational properties of RMS are expected
to vary between the different types of sources.

Early work on RMS
\citep{Pa66,ZR67,W76, BP81a, BP81b,LS82,Rif88}
 was restricted the Newtonian regime, where the diffusion approximation holds.   Unfortunately, the limited range of shock velocities 
that can be analyzed by employing the diffusion approximation renders its applicability to most high-energy transients of little relevance.
In the last decade there has been a growing interest in extending the analysis to the relativistic regime, in an attempt to identify observational diagnostics of these shocks, and in particular the early signal expected from shock breakout in various cosmic explosions, and the contribution of sub-photospheric shocks to prompt GRB emission.    An elaborated account of the astrophysical motivation is given in Section \ref{sec:Astro-Mot}. 

There are vast differences between relativistic and non-relativistic RMS, as described in  
\citet{LB08} and \citet{BKAW10}. 
In brief, in non-relativistic RMS the shock 
thickness is much larger than the photon mean 
free path and the  energy gain in a single collision
is small.    As a consequence, the diffusion approximation holds, which
considerably simplifies the analysis.   
In contrast, in relativistic and mildly-relativistic RMS the shock thickness is a few Thomson depths, the change in photon momentum 
in a single collision is large, the optical depth is highly anisotropic, owing to relativistic effects, and
pair production is important, even dominant in RRMS with cold upstream.  Thus, the analysis of RRMS 
is far more challenging and requires different methods. Monte-Carlo simulations is an optimal method for computations
of steady and slowly evolving shocks. The advantage of this method is its flexibility, that allows a systematic
investigation of a large region of the parameter space relevant to a variety of systems.

In this paper we present results of Monte-Carlo simulations of infinite planar RMS propagating in an unmagnetized plasma, 
in the regime where advection of photons by the upstream flow dominates over photon production inside and just downstream 
of the shock transition layer (photon rich shocks).  The regime of photon starved shocks will be presented in a follow up paper. 
In Section \ref{sec:Astro-Mot} we give an overview of the astrophysical motivation.  In Section \ref{sec:general} we derive some basic 
properties of a RRMS and compute analytically 
its structure. In Section \ref{sec:MC-code} we describe the code and the method of solutions.  The results are presented in Section \ref{result}, and the applications to 
GRBs in Section \ref{application}. We conclude in Section \ref{conclusion}.

\section{Astrophysical motivation}
\label{sec:Astro-Mot} 
This section briefly summarizes the astrophysical motivation for considering RMS.  
We focus on shock breakout in stellar explosion, including chocked GRB jets, and
photospheric GRB emission.

\subsection{Shocks generated by stellar explosions}

The first light that signals the death of a massive star is emitted upon emergence of the shock 
wave generated by the explosion at the surface of the star.
Prior to its breakout the shock propagates in the dense stellar envelope, and is 
mediated by the radiation trapped inside it.   The observational signature of the  breakout event depends on the shock 
velocity and the environmental conditions, thus, detection of the breakout signal and the subsequent emission can provide a wealth 
of information on the progenitor (e.g., mass, radius, mass loss prior to explosion, etc.) and on the explosion mechanism.   
Recent observational progress has already led to the discovery of a few shock breakout candidates, notably SN 2008D, and next generation
transient surveys promise to detect many more. 
In order to extract this information the structure of the RMS must be computed.

A shock that traverses the stellar envelope during a SN explosion
accelerates as it propagates down the declining density gradient near the stellar edge, and while the bulk of the material is always non-relativistic in SNe, the accelerating shock can bring, in some cases, a small fraction of the mass to mildly and even ultra-relativistic velocities. For a typical spherical explosion with an energy of $\sim 10^{51}$ erg this happens in compact stars with $R_* \lesssim R_\odot$, while for larger energies, and/or strongly collimated explosions, relativistic shocks are generated also in more extended progenitors
\citep{Tan01, NS12, Na15}.

\citet{BKAW10}
 obtained solutions of infinite planar RMS in the ultra-relativistic limit, under conditions anticipated in SN shock breakouts. 
These solutions are valid within the star where the shock width is much smaller than the scale over which the density vary significantly.
\citet{NS12}
 employed those solutions to show that a relativistic shock breakout from a stellar surface gives rise to a flash of 
gamma-rays with very distinctive properties.    Their analysis is applicable to progenitors in which the breakout is sudden, as in cases 
where transition to a collisionless shock occurs near the stellar surface,  but not to the gradual breakouts anticipated in situations wherein 
the progenitor is surrounded by a stellar wind thick enough to sustain the RMS after it emerges from the surface of the star.    
In the latter case, the gradual evolution of the shock during  the breakout phase can significantly alter the breakout signal. 
This case is of special interest since Wolf-Rayet stars, which are thought to be the progenitors of long GRBs, and are also compact 
enough to have relativistic shock breakout in extremely energetic SNe (such as SN 2002ap),  are known to drive strong stellar winds. 
Shock breakout from a stellar wind has been studied in the non-relativistic regime \citep{SN14a, SN14b}.  Analytic solutions for 
the structure of ultra-relativistic RMS with gradual photon leakage have also been found recently
\citep{GNL17}.  
We plan to carry out a comprehensive analysis of such shocks is the near future. 

\subsection{Implications for high-energy neutrino production in GRBs}

It has been proposed that the interaction of photons with protons accelerated to high energy in shocks during jet propagation through the stellar envelop may produce, for both GRB jets and slower jets that may be present in a larger fraction of core collapse SNe, bursts of $\sim1$~TeV neutrinos  \citep[e.g.,][]{MW01, RMW03}.   
However, in early models the fact that internal and collimation shocks that
are produced below the photosphere are mediated by radiation has been overlooked.    
As explained below, in such shocks particle acceleration is extremely inefficient.    This has dramatic implications for neutrino 
production in GRBs
\citep{LB08, MI13, GAM15}.
This problem can be avoided in 
ultra-long GRBs \citep{MI13}  and in low-luminosity GRBs
\citep{Na15, SMM16}, if indeed produced by choked GRB
jets, as in the unified picture proposed by
\citet{Na15}.  In the latter scenario, the progenitor star is ensheathed by an extended 
envelope that prevents jet breakout.  If the jet is chocked well above the photosphere, then internal shocks produced inside the jet 
are expected to be collisionless.   The photon density at the shock formation site may still be high enough to contribute the photo-pion 
opacity required for production of a detectable neutrino flux. 

Substantial magnetization of the flow may alter the above picture, because in this case 
formation of a strong collisionless subshock within the RMS occurs \citep{B17}.  
Whether particle acceleration is possible in mildly relativistic internal shocks with at relatively 
high magnetization is unclear at present \citep{SSA13}, but if it does then the problem of neutrino production in
GRBs should be reconsidered. 

\subsection{Sub-photospheric shocks in GRBs}

The composition and dissipation mechanisms of GRB jets are yet unresolved issues.
The conventional wisdom has been that those jets are powered by magnetic 
extraction of the rotational energy of a neutron star or an accreting black hole, and that the energy thereby extracted is transported 
outward in the form of Poynting flux, which on large enough scales is converted to kinetic energy flux.   An important question
concerning the prompt emission mechanism is whether the conversion of magnetic-to-kinetic energy occurs above or well below 
the photosphere
\citep[e.g.,][]{MU12, LBg13, BGL14}.
  Dissipation well 
below the photosphere is naturally  expected in case of a quasi-striped magnetic field
configuration
\citep{DS02, LG16}.
  Rapid dissipation of an ordered magnetic field may ensue
in a dense focusing nozzle via the growth of internal kink modes, as demonstrated recently by state-of-the-art numerical 
simulations 
\citep{BT16,SMG16}.
 In such circumstances, the GRB outflow is expected to be 
weakly magnetized when approaching the photosphere. 

If this is indeed the case, then further dissipation, that produces the observed prompt emission,
most likely involves internal and recollimation shocks in the weakly magnetized flow.   
Substantial dissipation is anticipated just below the 
photosphere for typical parameters
\citep{BML11, LMB09, MLB10, IMN15, B17, LBV17}.
  Various (circumstantial) indications of photospheric emission support
this view.   These include: (i) detection of a prominent thermal component in
several bursts, e.g., GRB090902B
\citep{Abd09, RAZ10}, and claimed evidence for
thermal emission in many others
\citep{PMR06, RP09};
 (ii) a hard spectrum below the peak that cannot be accounted
for by optically thin synchrotron emission;   (iii) evidence (though
controversial) for clustering of the peak energy around 1 MeV, that
is most naturally explained by photospheric emission;  In addition,
an attractive feature of sub-photospheric dissipation models is that
they can lead to the high radiative efficiency inferred from observations.

A large body of work on photospheric emission does not address the nature of the dissipation mechanism and
the issue of entropy generation.  Earlier work
\citep[e.g.,][]{PMR06, Bl13, Vetal13}
 attempted
to compute the evolution of the photon density below the photosphere, assuming dissipation by some unspecified mechanism.   
They generally find significant broadening of the seed spectrum if dissipation commences in sufficiently opaque regions
and proceeds through the photosphere. 
\citet{KL14} have reached a similar conclusion, demonstrating 
that multiple RMS can naturally generate a Band-like
spectrum.  More recent work
\citep{IMN15, L16, PL17}
 combines hydrodynamics (HD) and Monte-Carlo codes to compute the emitted spectrum.
In this technique, the output of the HD simulations is used as input for the Monte-Carlo radiative transfer calculations.  
These calculations illustrate that a Plank distribution, injected at a large optical depth, evolves into a 
Band-like spectrum owing to bulk Compton 
scattering on layers with sharp velocity shears, mainly 
associated with reconfinement shocks.   However, one must be cautious in applying those results, since the emitted
spectrum is sensitive to the width of the boundary shear layers
\citep{INO13}, which is unresolved in those simulations.
Furthermore, the radiative feedback on the shear layer is ignored.  Ultimately, the structure of those radiation mediated 
reconfinement  shocks needs to be resolved to check the validity of the results.

The Monte-Carlo simulations described in Section \ref{result} 
enable detailed calculations of the spectrum produced in a sub-photospheric GRB shock
prior to its breakout.   The implications for prompt GRB emission are discussed in
Section \ref{application}. 


\section{General considerations and summary of previous work}
\label{sec:general}

Consider an infinite planar shock propagating in an unmagnetized plasma at a velocity $\beta_u$ (henceforth measured in units of the speed of light $c$).
In the frame of the shock, the jump conditions read:
\begin{eqnarray}
n_{u}\gamma_u\beta_u=n_{d}\gamma_d\beta_d,\label{jmp-1}\\
w_u\gamma_u^2\beta_u^2+p_u=w_d\gamma_d^2\beta_d^2+p_d,\label{jmp-2}\\
w_u\gamma_u^2\beta_u=w_d\gamma_d^2\beta_d,\label{jmp-3}
\end{eqnarray}
where $n$ denotes the baryon density, $w$ the specific enthalpy, $\beta$ the fluid velocity with respect to the shock frame, and $\gamma=(1-\beta^2)^{-1/2}$ the Lorentz factor.   The  subscripts $u$ and $d$ refer to the upstream and downstream values of the fluid 
parameters, respectively.  Radiation dominance is established in the downstream plasma when the shock velocity satisfies
\begin{eqnarray}
\beta_{u} > 4\times10^{-5} n_{u 15},
\end{eqnarray}
where $n_{u 15} = n_u / 10^{15}~{\rm cm^{-3}}$.
At velocities well in excess of this value the shock becomes 
radiation mediated
\citep{W76, BKAW10}.
This readily implies that under conditions anticipated in essentially all compact astrophysical systems, 
relativistic and mildly relativistic shocks that form in opaque regions are mediated by radiation.

It is insightful to compare different scales that govern microphysical interactions.   
The width of the RMS transition layer is typically on the order of the photon diffusion length,
\begin{eqnarray}
l^\prime_s\sim (n_u \sigma_T\beta_u)^{-1}\simeq10^9(\beta_u n_{u 15})^{-1}\quad {\rm cm},
\end{eqnarray}
here measured in the shock frame.   As shown below, it can be substantially smaller when pair production is important, but
by no more than  
three orders of magnitudes.
    The skin depth is 
\begin{eqnarray}
\delta \sim c/\omega_p\simeq 1\, n_{15}^{-1/2} \quad {\rm cm}, 
\end{eqnarray}
where $\omega_p$ is the plasma frequency,  and the gyroradius of relativistic protons of energy $\epsilon_p$ is,  
\begin{eqnarray}
r_L\sim3\left(\frac{\epsilon_p}{m_pc^2}\right) \left(\frac{B}{10^6\ {\rm G}}\right)^{-1}\quad  {\rm cm}. 
\end{eqnarray}
Evidently, under typical conditions kinetic processes are expected to play no role in RMS, with the exception of
 subshocks that form within the shock transition layer in certain cases (see below). In particular, the 
extremely small ratio, $r_L/l_s^\prime \sim 10^{-8}$, implies that particle acceleration is unlikely in RMS. 
Kinetic processes become important, of course, during the transition phase from RMS to collisionless shocks.

The properties of the downstream flow of RMS depend on the parameters of the upstream flow, and in particular 
its velocity, magnetization, specific entropy and optical depth.  The following discussion elucidates the 
effect of each of these parameters.

\subsection{Photon sources}
The main sources of photons inside and just downstream of the shock transition layer are photon advection by the upstream flow,
and photon generation by Bremsstrahlung and double Compton emission.  Each process dominates in a different regime.   In what follows
"photon rich" shocks refer to RMS in which photon generation is negligible and "photon starved" shocks to RMS in which photon advection
is negligible.   The advected radiation field is henceforth characterized by two parameters: the photon-to-baryon density ratio far
upstream, $\tilde{n}=n_{\gamma u}/n_u$, and the fraction $\xi_\gamma$ of the total energy which is carried by radiation in the unshocked flow. 
Under the conditions anticipated in RRMS the thermal energy of the upstream plasma is  negligible, so that to a good 
approximation one has
\begin{eqnarray}
\xi_\gamma=\frac{\gamma_u\,e_{\gamma u}}{(\gamma_u-1)m_pc^2n_u+\gamma_u\,e_{\gamma u}},
\end{eqnarray}
where $e_\gamma=3p_\gamma$ denotes the energy density of the radiation field.  

The specific photon generation rate by thermal Bremsstrahlung emission can be expressed as  
\begin{eqnarray}
\dot{n}_{ff}&=&\frac{2^{3/2}\alpha_f\sigma_Tcn_{i}^2}{\sqrt{3\pi}\,\Theta^{1/2}}\{(1+2x_+)\Lambda_{ep}\\ \nonumber
&+&[x_+^2+(1+x_+)^2]\Lambda_{ee}+x_+(1+x_+)\Lambda_{+-}\},
\label{rate-ff}
\end{eqnarray}
here $n_i$ is the ion density, $x_+=n_+/n_i$ is the pair multiplicity, specifically the number of positrons per ion, 
$\Theta=kT/m_ec^2$ denotes the plasma temperature in units of the 
electron mass, and $\alpha_f$ is the fine structure constant.  The total number of electrons is dictated by charge neutrality, $n_e=(1+x_+)n_i$.
The terms labeled $ep$ and $+-$ account for the contributions of $e^\pm p$ and $e^+ e^+$ encounters, respectively, and
the term $ee$ for the contribution of $e^- e^-$ and $e^+ e^+$ encounters.    The quantities $\Lambda_{ep}$, $\Lambda_{ee}$ and $\Lambda_{+-}$
are functions of the temperature $\Theta$, and are given explicitly in \cite{Ski95}.

The rate of double Compton (DC) emission is
\begin{eqnarray}
\dot{n}_{DC}=\frac{16}{\pi}\alpha_f\sigma_Tcn_{e}n_{\gamma }\Theta^{2}\Lambda_{DC},\label{rate-DC}
\end{eqnarray}
with $\Lambda_{DC}$  given in
\citet{BML11}.
  As the ratio of the two rates is $\dot{n}_{DC}/\dot{n}_{ff}\sim (n_\gamma/n_e)\Theta^{5/2}$,
it is readily seen that DC emission is only important in regions where the photon density largely exceeds the total lepton density 
$n_\gamma \gtrsim n_e\,\Theta^{-5/2}$.    Such conditions prevail in the near downstream of sufficiently photon rich shocks
\citep{LE12}.
In photon starved shocks, where $n_\gamma\simeq n_e$ and $\Theta\simeq 0.2$, DC emission is negligible.  As shown in Section \ref{result}, this is also true for photon rich shocks with small enough $\tilde{n}$, in which the density of pairs produced by nonthermal photons  becomes substantial, $x_+ \gg 1$.

We now give a crude estimate of the advected photon density above which the shock is expected to be photon rich and below
which it is photon starved
\citep[see also][]{BML11}.
We note first that photons which are produced downstream can 
diffuse back and interact with the upstream 
flow provided they were generated roughly within one diffusion length, $L_d=(\sigma_T n_{l}\beta_d)^{-1}$, from the shock transition layer,
where $n_l=(1+2x_+)n_d$ is the total lepton density in the immediate post shock region.    The density of these photons is 
$\delta n_\gamma=\dot{n}_\gamma L_d/c\beta_d$.    Since for marginally rich shocks photon generation is dominated by Bremsstrahlung, we
have to a good approximation
\begin{eqnarray}
&&\frac{\delta n_\gamma}{n_{\gamma d}}\simeq \frac{2^{3/2}\alpha_f n_{d}}{\sqrt{3\pi}\Theta_d^{1/2}\beta_d^2n_{\gamma d}}\\
&\times&\left\{\Lambda_{ep}+\frac{[x_+^2+(1+x_+)^2]\Lambda_{ee}}{(1+2x_+)}+\frac{x_+(1+x_+)\Lambda_{+-}}{(1+2x_+)}\right\},\nonumber
\end{eqnarray}
where Equation (\ref{rate-ff}) has been used with $n_i=n_d$.  Assuming that the shock is photon rich, we employ Equations (\ref{phot_cont}) 
and  (\ref{shock_rich_temp}) below to get
\begin{eqnarray}\label{delta_n/n_gamma}
&&\frac{\delta n_\gamma}{n_{\gamma d}}\simeq\frac{3\times10^{-3}}{\sqrt{\tilde{n}\gamma_u\beta_u}}\\
&\times&\left\{\Lambda_{ep}+\frac{[x_+^2+(1+x_+)^2]\Lambda_{ee}}{(1+2x_+)}+\frac{x_+(1+x_+)\Lambda_{+-}}{(1+2x_+)}\right\}. \nonumber
\end{eqnarray}
Typically, the terms $\Lambda_{12}$ lie in the range  $10< \Lambda_{12} < 20$, and since $\tilde{n}>10^3$ for photon rich shocks 
(see Equation (\ref{ncrit}) below) it is 
evident that photon generation is important only when pair loading is substantial ($x_+ \gg 1$).   Adopting for illustration 
$\Lambda_{ee}+\Lambda_{+-}/2=30$ we estimate that photon generation will dominate over photon advection when 
$x_+>10\sqrt{\tilde{n}\,\gamma_u\beta_u}$.

\begin{figure}
\centering
\includegraphics[width=8cm]{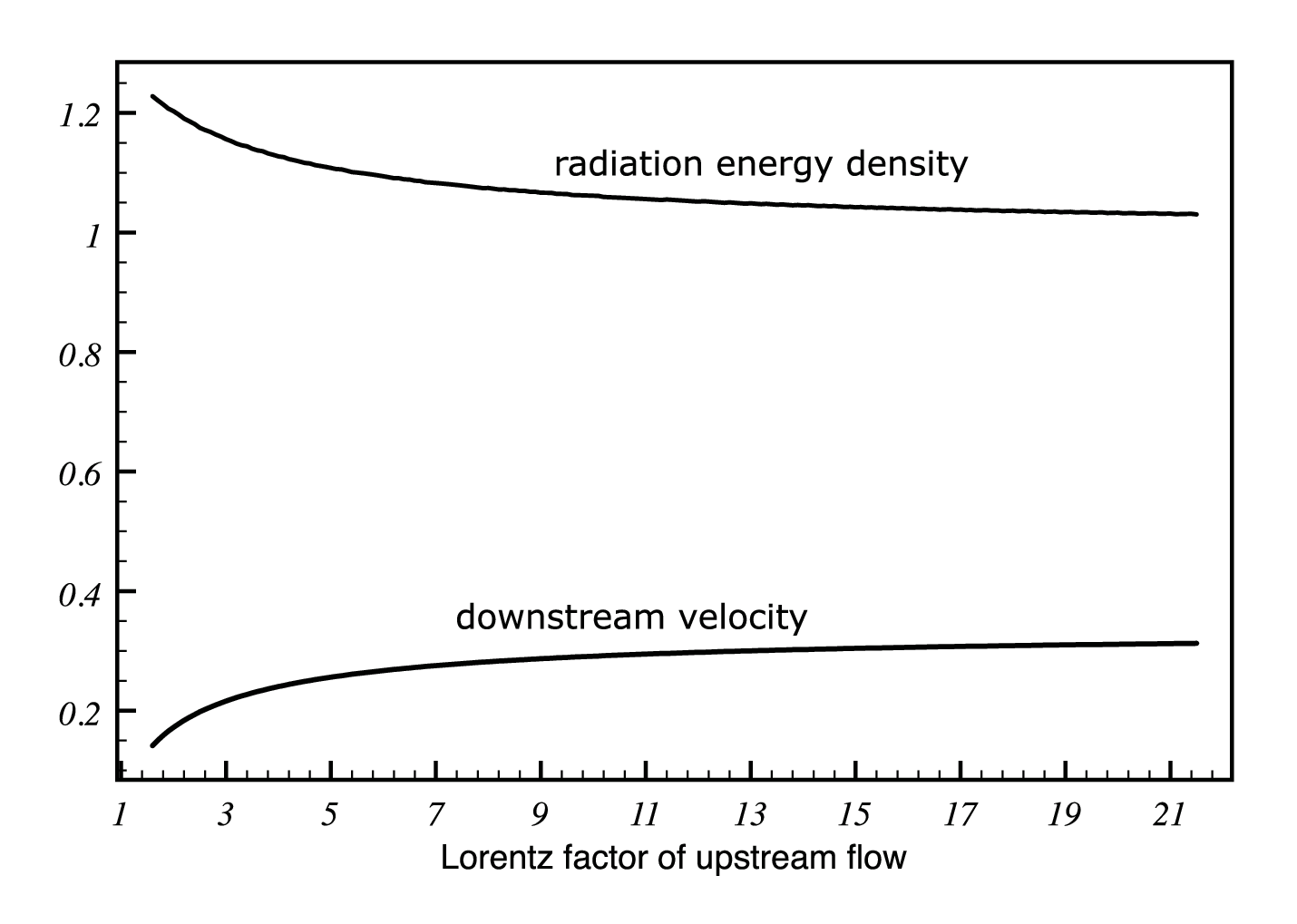}
\caption{\label{fig:f1} Velocity, $\beta_d$, and normalized radiation energy density, $e_{\gamma d}/(2m_pc^2n_u\gamma_u^2\beta^2_u)$, 
of the downstream flow as functions of the upstream Lorentz factor $\gamma_u$.}
 \end{figure}

\subsection{Photon rich regime}
\label{sec:photon_rich}
As shown in Section \ref{result}, in photon rich shocks the density of pairs produced by nonthermal photons is typically much smaller than 
the density of the radiation, and while under certain conditions the pairs can dominate the opacity inside the shock and affect 
its structure, they contribute very little to the total  energy budget downstream. 
The specific enthalpy is then well approximated by $w=n m_pc^2+4p_{\gamma}$, where $p_\gamma$ denotes the radiation pressure.
Adopting the latter equation of state, the jump conditions, Equations (\ref{jmp-1})-(\ref{jmp-3}), can be solved to yield the parameters of the downstream flow. 
Solutions for the downstream velocity, $\beta_d$, and the specific radiation energy, $e_{\gamma d}=3p_{\gamma d}$,  are exhibited in Fig. \ref{fig:f1} in the limit $\xi_\gamma \ll 1$.   The radiation energy in Fig. \ref{fig:f1} is normalized to the value obtained in the ultra-relativistic limit (i.e., for $\beta_d=1/3$):  
\begin{eqnarray}
e_{\gamma d}=2 n_{u}\gamma_u^2\beta_u^2m_pc^2.\label{erR}
\end{eqnarray}
As seen, this asymptotic value is a good approximation also at mild Lorentz factors, and is adopted for illustration 
in the following discussion. 

The downstream region of a RRMS is inherently non-uniform, because the thermalization length 
over which the plasma reaches full thermodynamic equilibrium is larger than the width
of the shock transition layer. 
However, for typical astrophysical parameters, the thermalization length exceeds the 
shock width by several orders of magnitudes
\citep{LE12}, 
 so that for any practical
purpose photon generation in the downstream plasma can be ignored.  
This readily implies that to a good approximation the photon number is conserved across the 
shock transition layer, whereby
\begin{eqnarray}
n_{\gamma d}\gamma_d\beta_d=n_{\gamma u}\gamma_u\beta_u.
\label{phot_cont}
\end{eqnarray}
Combining with  Equation (\ref{jmp-1}), one finds $\tilde{n}=n_{\gamma u}/n_{u}=n_{\gamma d}/n_{d}$.
The temperature can be computed using Equations (\ref{erR}) and (\ref{phot_cont}), yielding
\begin{eqnarray}
\Theta_d =\frac{e_{\gamma d}}{3n_{\gamma d}\,m_ec^2}=\frac{2m_p}{3\,m_e}\frac{(\gamma_u\beta_u)(\gamma_d\beta_d)}{\tilde{n}}
\simeq 430\,\frac{\gamma_u\beta_u}{\tilde{n}},
\label{shock_rich_temp}
\end{eqnarray}
where $\beta_d=1/3$ was adopted to obtain the numerical factor in the rightmost term. 
Thus, $\Theta_d \ll 1$ as long as $\tilde{n} \gg 430\gamma_u\beta_u$.    

Next, we estimate the minimum value of $\tilde{n}$ required in order that counter-streaming photons will be able to
decelerate the upstream flow.   Let $\eta$ denotes the fraction of downstream photons that 
propagate towards the upstream.   The average energy each counter-streaming photon can extract in a single collision 
is at most $\gamma_um_ec^2$.   Thus, the number of downstream photons required to decelerate the upstream flow satisfies
$\gamma_d\,n_{\gamma d}>\eta^{-1} (m_p/m_e)\gamma_u\,n_{u}$ (assuming $\xi_\gamma \ll 1$).  By employing Equation (\ref{phot_cont}) we find that the shock can be mediated by the advected 
photons provided the photon-to-baryon number ratio far upstream satisfies
\begin{eqnarray}
\tilde{n}>\tilde{n}_{crt}\equiv \frac{m_p}{m_e}\frac{\beta_d}{\eta \beta_u}\simeq \frac{m_p}{m_e},
\label{ncrit}
\end{eqnarray}
adopting $\beta_d/\eta =1$, which is roughly the value obtained from the Monte-Carlo simulations.
At the critical number density the average photon energy, $3 k T_d \simeq 2\eta\,m_ec^2 \gamma_u\beta_u$, 
is in excess of the electron mass.  Under this condition a vigorous pair production is expected to ensue inside 
and just downstream of the shock, that will significantly enhance photon generation, thereby 
regulating the downstream temperature. 

\subsubsection{Analytic shock profile}
\label{sec:analyt}

Let  $n_{\gamma\rightarrow u}$ denotes the density of photons moving in upstream direction (i.e., from downstream to the upstream), as measured
in the shock frame, and $n$, $n_e$, $n_\pm$ the proper densities of baryons, electrons and $e^\pm$ pairs, respectively. 
We suppose that the flow moves along the $z$-axis, chosen such that  its positive direction is towards the upstream.   The 
change in the photon density is governed by the equation
\begin{eqnarray}
\frac{dn_{\gamma\rightarrow u}}{dz}=-\sigma_{KN}\gamma (n_e+n_\pm) n_{\gamma\rightarrow u}.
\end{eqnarray}
For sufficiently photon-rich shocks the scattering of bulk photons is in 
the Thomson regime. In terms of the optical depth, $d\tau_{*} = \sigma_{T}\,(n_e+n_\pm)\gamma dz$, and the energy density of 
the counter streaming photons, $u_{\gamma\rightarrow u}=<\epsilon_\gamma>n_{\gamma\rightarrow u}$, one then has
\begin{eqnarray}
\frac{du_{\gamma\rightarrow u}}{d\tau_{*}}=-u_{\gamma\rightarrow u}.\label{dugamdtau}
\end{eqnarray}
The total inverse Compton power emitted by a single electron (positron) inside the shock is approximately 
\begin{eqnarray}
P_{Comp}= \kappa_\gamma c\sigma_T (\gamma\beta)^2\,u_{\gamma\rightarrow u},\label{Pcomp}
\end{eqnarray}
where the pre-factor $\kappa_\gamma$ ranges from $4/3$ for isotropic radiation to $4$ for completely beamed radiation.
For simplicity, we shall assume that it is constant throughout the shock.  

\begin{figure}
\centering
\includegraphics[width=8cm]{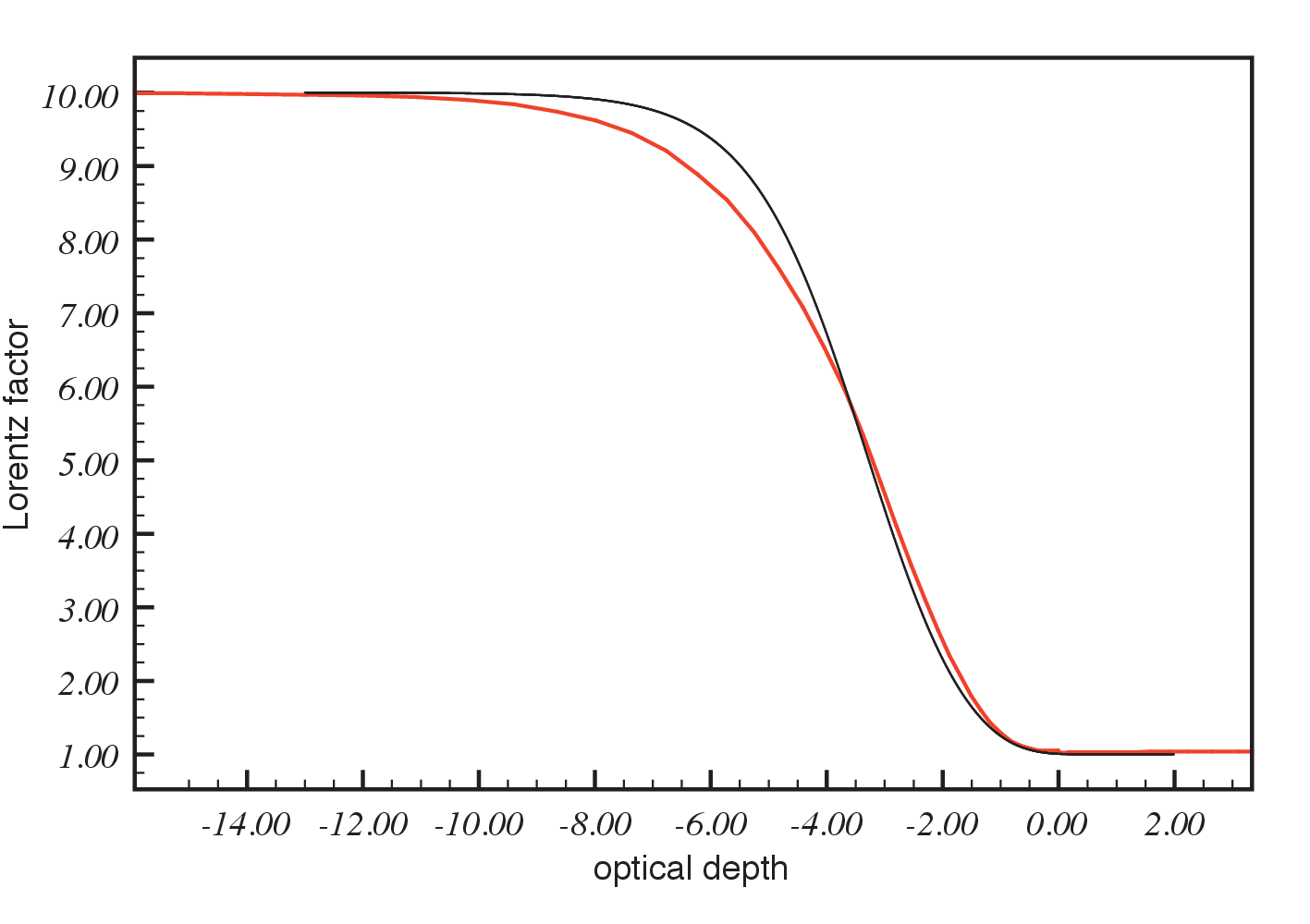}
\caption{\label{fig:f2} The solid black line delineates the solution given by Equation (\ref{analyt-prof-gam}) with $\gamma_u=10$ and
$\kappa_\gamma\epsilon=0.2$.  The red line is the shock profile obtained from the Monte-Carlo simulation.}
 \end{figure}

Neglecting the internal energy relative to baryon rest mass energy, the energy flux of the plasma can be expressed as 
\begin{eqnarray}
T_b^{0z}=m_pc^2n \gamma^2 =J \gamma,
\end{eqnarray}
in terms of the conserved mass flux,
\begin{eqnarray}
J=m_pc^2\,n \gamma.
\end{eqnarray}
Using Equation (\ref{Pcomp}) we obtain
\begin{eqnarray}
\frac{d\,T_b^{0x}}{dz}&=&\gamma(n_e+n_\pm) c^{-1} P_{comp} \\
&=&\kappa_\gamma\sigma_T\,\gamma(n_e+n_\pm)(\gamma^2-1) u_{\gamma\rightarrow u},\nonumber
\end{eqnarray}
or 
\begin{eqnarray}
J \frac{d\,\gamma}{d\tau_{*}}=\kappa_\gamma (\gamma^2-1) u_{\gamma\rightarrow u}.\label{dgdtau}
\end{eqnarray}
The boundary condition is $\gamma(\tau_{*}\rightarrow\infty)=\gamma_u$.
Denoting $\alpha=\kappa_\gamma u_{\gamma\rightarrow u}(\tau_{*}=0)/J$, and 
\begin{eqnarray}
\eta(\tau_{*})=\ln\left(\frac{\gamma_u+1}{\gamma_u-1}\right)+2\alpha\,e^{-\tau_{*}}
\label{analyt-shock-prof}
\end{eqnarray}
the solution of Equations (\ref{dugamdtau}) and (\ref{dgdtau}) reads:
\begin{eqnarray}
\gamma(\tau_{*})=\frac{e^\eta+1}{e^\eta-1}.
\label{analyt-prof-gam}
\end{eqnarray}
From the jump conditions we have $\alpha=2\kappa_\gamma\epsilon \gamma_u$, where 
$\epsilon=u_{\gamma\rightarrow u}/u_{\gamma d}$ is roughly the 
fraction of downstream photons that propagate backwards.  
The black solid line in  Fig. \ref{fig:f2} shows the shock profile obtained for $\kappa_\gamma \epsilon=0.2$.
The red line is the result of the full simulation.

\subsection{Photon starved regime}
At $\tilde{n}<\tilde{n}_{crt}$ photon advection is negligible, and the prime source of photons inside and downstream of the 
shock transition layer is free-free emission.  For sufficiently fast shocks, $\beta_u>0.6$, a pair production equilibrium 
is established downstream, keeping the temperature at $\Theta_d\lesssim m_ec^2/3$ \citep{KBW10}. 
In relativistic shocks, $\gamma_u \gg 1$, the downstream photon density is obtained from 
Equation (\ref{erR}) and the relation $e_{\gamma d}\simeq m_ec^2 n_{\gamma d}$': 
\begin{eqnarray}
n_{\gamma d}\simeq 2\frac{m_p}{m_e}n_u\,\gamma_u^2.
\label{pht-starv-nd}
\end{eqnarray}
This, in turn, implies that the transition from photon rich to photon starved shocks should in fact
occur at $\tilde{n}\simeq \tilde{n}_{crt}\gamma_u$.

Because the average energy of downstream photons is roughly $m_ec^2$, scattering of counter streaming photons
off electrons (positrons) in the shock transition layer is in the Klein-Nishina regime.   As a consequence, the temperature 
at any position $z$ inside the shock is expected to be comparable to the local bulk energy of the leptons, 
specifically $\Theta(z)\simeq \gamma(z)$.   Indeed, this result has been verified by detailed simulations \citep{BKAW10}. 
When this relation is adopted it is possible to compute the shock structure analytically in the limit 
$\gamma_u \gg 1$ \citep{NS12, GNL17}.
    The analysis indicates that the shock width increases with Lorentz factor according to 
$$
l_s^\prime\simeq10^{-2}\gamma_u^2\,(\sigma_T n_u)^{-1},
$$
where the numerical coefficient is somewhat arbitrary
\citep{NS12}. 
  This scaling stems from Klein-Nishina effects, and is different than
the scaling obtained for photon rich shocks. 
The photon spectrum exhibits (in the shock frame) a peak at $h\nu_{peak}\simeq m_ec^2$, with a broad (a rough
power law) extension up to energies $> \gamma_u m_ec^2$ \citep{BKAW10}.

\subsection{Effect of magnetic fields}

Substantial magnetization of the upstream flow can significantly alter the shock profile and 
emission. A prominent feature of such shocks is the formation of a relatively strong subshock 
\citep{B17}.   The results exhibited in Section \ref{result}
 indicate that subshocks
form also in unmagnetized shocks under  certain conditions (see Figs. \ref{g2e-2n5sub}, \ref{g4e-2n5sub} and
\ref{g10n5sub}), but those are generally 
weak and have little effect on the overall shock structure and emission, with the exception of the 
breakout transition, where photon leakage 
becomes important.  In magnetized shocks formation of strong subshocks is anticipated even in regions of large optical depth,
which can considerably alter the energy distribution of particles in the shock if particle acceleration at the collisionless subshock ensues. 
The net amount of energy that can be transferred to the nonthermal particles depends primarily on the relative strength of the subshock, and
needs to be quantified.   The formation of a strong subshock in RRMS may have profound implications for emission of sub-photospheric 
GRB shocks, as well as for neutrino production in chocked jets, as described above.

\subsection{Finite shocks and breakout}
 
The structures computed in \citet{BKAW10} and in Section \ref{result} are applicable to RRMS propagating in a medium of infinite optical depth. 
In cases where the shock propagates in a medium of gradually decreasing optical depth, e.g., stellar wind, it will eventually reach 
a point at which the radiation trapped inside it starts escaping to infinity.  The leakage of radiation leads to a steepening of the shock, 
at least in photon starved RRMS.   Nonetheless, the shock remains radiation mediated also at radii at which the optical thickness 
of the medium ahead of the shock is much smaller than unity,  owing to self-generation of its opacity through accelerated pair creation.  Breakout 
occurs  when the Thomson thickness of the unshocked medium becomes smaller than $(m_e/m_p)\gamma_u$, provided $\gamma_u>1$ , 
or else in the Newtonian regime \citep{GNL17}.
How this affects the emitted spectrum is yet to be explored.

A similar process may take place also during the breakout of a photon rich shock, since photon escape from the medium 
ahead of the shock (i.e., the upstream plasma) is expected to lead to the gradual decline of $\tilde{n}$ over time.  If
pair creation and photon generation occur over a time shorter than the breakout time, then 
a transition from photon rich to photon starved shock is expected prior to breakout, at least in cases where the shock remains relativistic in the frame
of the unshocked medium.  Otherwise the shock will evolve in some complex manner.  
In any case,  the spectrum emitted during the breakout phase may be altered.

\section{Monte-Carlo simulations of RRMS}
\label{sec:MC-code}

\subsection{Description of the Monte-Carlo code}
The Monte-Carlo code used by us enables computations of mildly relativistic and fully
relativistic radiation mediated shocks in a planar geometry, for
arbitrary upstream conditions.  It incorporates an
energy-momentum solver routine that allows adjustments of the shock
profile in each iterative step.  A photon source is placed sufficiently
far upstream, and is tuned to account for the assumed
photon density advected by the upstream flow.  
In each run, an initial shock profile is imposed (usually some parametrized
analytic function that satisfies the shock jump conditions) during
some initial stage at which photons that were injected upstream and
crossed the shock are accumulated downstream. Once the photon
density downstream reaches a level that ensures stability of the
system, the energy-momentum solver is switched on, and the shock
profile is allowed to change iteratively, until a steady state is
reached whereby energy and momentum conservation of the entire
system of particles (i.e., photons, baryons and electron-positron
pairs) is satisfied at every grid point.  Choosing the initial shock
profile such that it satisfies the jump conditions in the frame of
the simulation box guarantees that the final shock solution is
stationary in this frame (otherwise it propagates across the box
accordingly).   Since the jump conditions depend only on the
parameters of the upstream flow, they can be determined a-priori for
any given set of upstream conditions.

The present version of the code includes the following radiation processes:
Compton scattering, pair production  and annihilation, and energy-momentum 
exchange with the bulk plasma. Its applicability is therefore restricted to photon rich shocks.
We are currently  in the process of incorporating also
internal photon sources, specifically relativistic Bremsstrahlung
and double Compton scattering, that would allow simulations of
shocks for any upstream conditions, and in particular photon starved shocks. 
Magnetic fields can also be included upon a simple modification of the energy-momentum solver,
and is planned for a future work.

In developing the prescription for the iteration in our code, we 
mimic the method used in the context of
 relativistic cosmic ray modified shocks \citep{EWB13}.
The difference is that, while we track photons,
they track cosmic rays using Monte-Carlo technique and 
evaluate the feedback on the bulk shock profile.

\subsection{Numerical setup}
\label{Numsetup}
 In our calculations, the input parameters are the following quantities at the
 far upstream region: (i) the  photon-to-baryon inertia ratio,
 $\xi_{u *} \equiv e_{\gamma u} / (n_u m_p c^2)$,  (ii) the photon-to-baryon number ratio,  $\tilde{n} \equiv n_{\gamma u}/n_u$, and (iii) the 
 bulk Lorentz factor of the upstream flow with respect to the shock frame, $\gamma_{u}$.
 Once these parameters are determined, all the physical quantities 
 at the far upstream region
 are specified under the assumption that the radiation and the plasma (protons and electrons) have identical temperature, $T_u$.
 We further assume that  the photons and the plasma constituents in the upstream region have 
 Wien and Maxwell distributions, respectively.
 For given upstream conditions, we derive the corresponding steady shock solution using the iterative scheme described in the previous section.\footnote{
Note that the shock structure 
 can be determined without specifying the absolute value
of the baryon density (or, equivalently, photon number density) 
 when  expressed as a function of optical depth.
The obtained solution is scale-free in which the number densities of photon and pairs
are only described  in terms of the ratio to that of the baryons.
The determination of the baryon number density gives the absolute values of these quantities
as well as the physical spatial scale.
This is valid as long as effects such as free-free absorption that break the scalability are ignored.}

In each iterative step, we solve the radiation transfer using Monte-Carlo method
under a given plasma profile, and evaluate the 
energy-momentum exchange between the photon and plasma.
We continue the iteration until the deviation of the total energy-momentum flux
at every grid point from that of the steady state value becomes small.
 In the calculations presented in this paper, the errors in 
 the conservation of momentum and energy fluxes
 after the iteration are mostly within a few \% throughout the entire structure ($< 15$ \% at most).
The total energy and momentum fluxes at each grid point are evaluated as
\begin{eqnarray}
F_{m}  =  \gamma^2 (\rho_{pl} c^2 + e_{pl} + p_{pl}) \beta^2 + p_{pl} + F_{m,\gamma},
\end{eqnarray}
and
\begin{eqnarray}
F_{e}  =  \gamma^2 (\rho_{pl} c^2 + e_{pl} + p_{pl}) \beta + F_{e,\gamma},
\end{eqnarray}
respectively.
Here $\rho_{\rm pl} = n m_p + (n + n_{\pm}) m_e$, $e_{pl} = 3/2 n k T + 3/2 f(T) (n + n_{\pm}) k T$, $p_{pl} = (2n + n_{\pm}) k T$ are the 
rest mass density, internal energy density and pressure of the plasma, respectively,
where $n_{\pm}$ is the number density of the created electron-positron pairs and 
$f(T) = {\rm tanh}[({\rm ln}\Theta + 0.3) / 1.93] + 3/2$ is an analytical function of temperature defined in 
 \citet{BKAW10}, obtained from a fit to the exact  equation of state of pairs at an arbitrary temperature ($f=1$  for $\Theta \ll 1$ and $f\approx 2$ for $\Theta \gg 1$).  
It is assumed that the protons and pairs have identical local temperature at every grid point.
The last terms in the above equations, $F_{m, \gamma}$ and $F_{e, \gamma}$, denote the momentum and energy fluxes
of radiation that are 
 directly computed by summing up the contributions of individual photon packets tracked in the Monte-Carlo simulation.
The steady state values of momentum, $F_{m,u}$, and energy fluxes, $F_{e,u}$, are evaluated by substituting 
 the enthalpy $w_u = n_u  (m_p + m_e) c^2 + [(7/2 + 3/2 f(T_u))n + 4 n_{\gamma u}] k T_u
 $ and pressure $p_u = (2n + n_{\gamma}) k T_u$ in the left side terms of Equations (\ref{jmp-2}) and (\ref{jmp-3}).

 At first,  the above iteration is performed under the assumption that a subshock
 is absent in the system. 
 If it converges to steady flow, we simply employ the solution 
 and consider that the  shock dissipation
 is solely  due to photon plasma interaction.
 On the other hand, when we find that the flow does not reach
 the steady state  under the assumption (error in energy-momentum flux is
 larger than $\sim 20 $ \%),
 we introduce a subshock in the system.
 The subshock is treated as a
 discontinuity in the plasma profile
 which satisfies the Rankin-Hugoniot conditions
 under the assumption that bulk plasma is isolated from the radiation.
 This setup is justified due to the fact that,
 since the plasma scale is much shorter
 than that of the photon mean free path,
 photons cannot feel the continuous change in the transition 
 layer of shock formed via plasma interactions.
 Once we introduce the subshock in the system, 
 we also vary the immediate upstream velocity in each iterative steps
 and continue the computation until it approaches to steady solution.

Regarding the microphysical processes, Compton scattering is evaluated using the full Klein-Nishina cross section. It is noted that,  in each scattering, bulk motion as well as thermal motion of the pairs are properly taken into account, under the assumption that the pairs have a Maxwellian distribution at the local temperature.
The rate of pair production is calculated based on the local photon distribution using the cross section given in \citet{GS67}. 
The pair annihilation rate is computed as a function of the local number density  and temperature.
Here we use the same analytical function employed in \citet{BKAW10} which is based on the formula given by \citet{S82}.
As for the spectra of photons generated via the pair annihilation process, we employ an analytical formula derived in \citet{SLP96} which is given as a function of temperature.   The details of the processes incorporated in our code are summarized in the appendix.

In this study we systematically explore the properties of RRMS, with a particular focus on the role of
the three parameters defined above.
We performed 15 model calculations that cover a wide range of parameters ($10^{-2} \leq \xi_{u *} \leq 10$, $10^3 \leq  \tilde{n} \leq 10^5 $, and $2 \leq \gamma_u \leq 10$).
Table \ref{tab1} summarizes the imposed values for each calculation.
The total number of injected photon packets varies among the models, but is typically in the 
range $N_{\rm pack} \sim 10^8 - 10^9$, which is sufficiently large to avoid significant statistical errors.





















\begin{table}
\begin{center}
\caption{Shock Parameters. \label{tab1} Column (1) shows the names of  the models.
 Columns (2), (3) and (4) display, respectively,
 the bulk Lorentz factor, 
  photon-to-baryon inertia ratio,
 and  photon-to-baryon number ratio at far upstream.}
\begin{tabular}{lllll}
\hline\hline

~Model  & $\gamma_u$ 
       & $\xi_{u*}$
       & $\tilde{n}$  \\

 ~~~~(1) & (2)
       & (3)
       & (4) \\
\hline

  g2e1n5 &  
           &  10 
	   &  \\ 

  g2e0n5 & 2 
           &  1 
	   & $10^5$ \\ 

  g2e-1n5 &  
           &  $10^{-1}$
	   & \\ 

  g2e-2n5 &  
           &  $10^{-2}$ 
	   &  \\ 

\hline

  g2e0n4 &  
           &  1 
	   &  \\ 

  g2e-1n4 & 2 
           &  $10^{-1}$ 
	   & $10^4$ \\ 

  g2e-2n4 &  
           &  $10^{-2}$ 
	   &  \\

\hline

  g2e-1n3 & 2 
           &  $10^{-1}$ 
	   & $10^3$ \\ 

  g2e-2n3 &  
           &  $10^{-2}$ 
	   &  \\

\hline

  g4e0n5 &  
           &  1 
	   &  \\ 

  g4e-1n5 & 4 
           & $10^{-1}$ 
	   & $10^5$ \\ 

  g4e-2n5 &  
           & $10^{-2}$ 
	   &  \\ 

\hline

  g10e0n5 &  
           &  1 
	   &  \\ 

  g10e-1n5 & 10 
           &  $10^{-1}$ 
	   & $10^5$ \\ 

  g10e-2n5 &  
           & $10^{-2}$ 
	   &  \\ 
\hline
\end{tabular}
\end{center}
\end{table}

\section{Results}
\label{result}

 In this section, we demonstrate how the different  parameters  affect
 the shock properties.
 Hereafter we refer to the models with $\gamma_u = 2$ and $\tilde{n} = 10^5$ 
 as fiducial cases (g2e1n5, g2e0n5, g2e-1n5 and g2e-2n5), since
 such  conditions are likely to prevail in sub-photospheric GRB shocks.
 Note that the cases $\xi_{u *} \geq 1$ and  $\xi_{u *} < 1$
 correspond to shocks formed below and above the saturation radius, respectively,
 in the context of the fireball model.

\subsection{Dependence on  $\xi_{u*}$}
\label{fiducial}

\begin{figure*}
\begin{center}
\includegraphics[width=16cm,keepaspectratio]{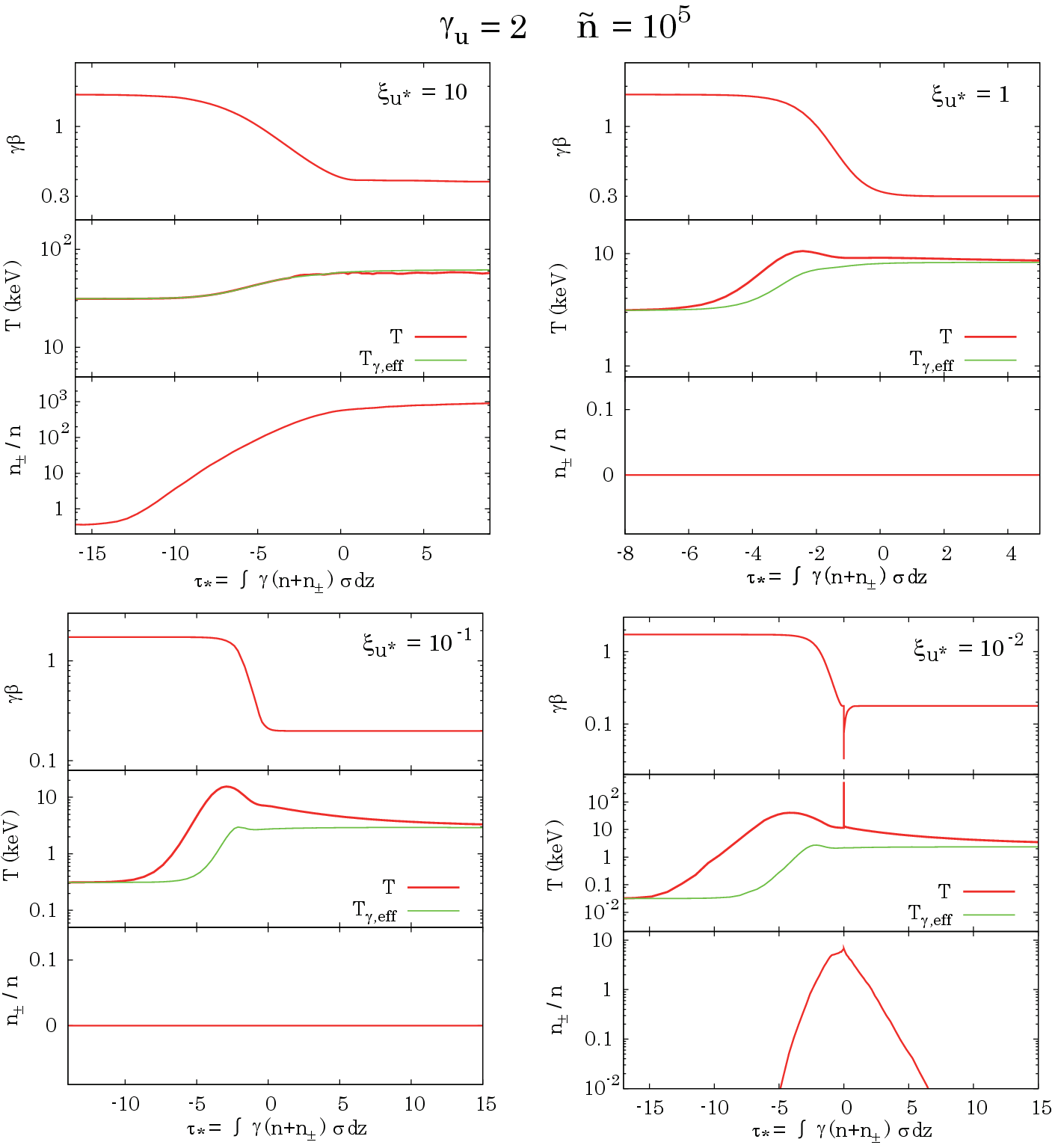}
\end{center}
\caption{The overall shock structure for models g2e1n5 ({\it top left}), g2e0n5 ({\it top right}), g2e-1n5 ({\it bottom left}) and g2e-2n5 ({\it bottom right}). In each panel, from top to bottom, we display the  4-velocity $\gamma \beta$, the plasma
 temperature $T$, the effective radiation temperature $T_{\gamma, {\rm eff}}$,
 and  the pair -to- baryon density ratio $n_{\pm}/n$, as a function of optical depth $\tau_{*}$.
Note the difference in the scaling of the horizontal and vertical axes in the different models.
}
\label{g2n5dist}
\end{figure*}

\begin{figure*}
\begin{center}
\includegraphics[width=14cm,keepaspectratio]{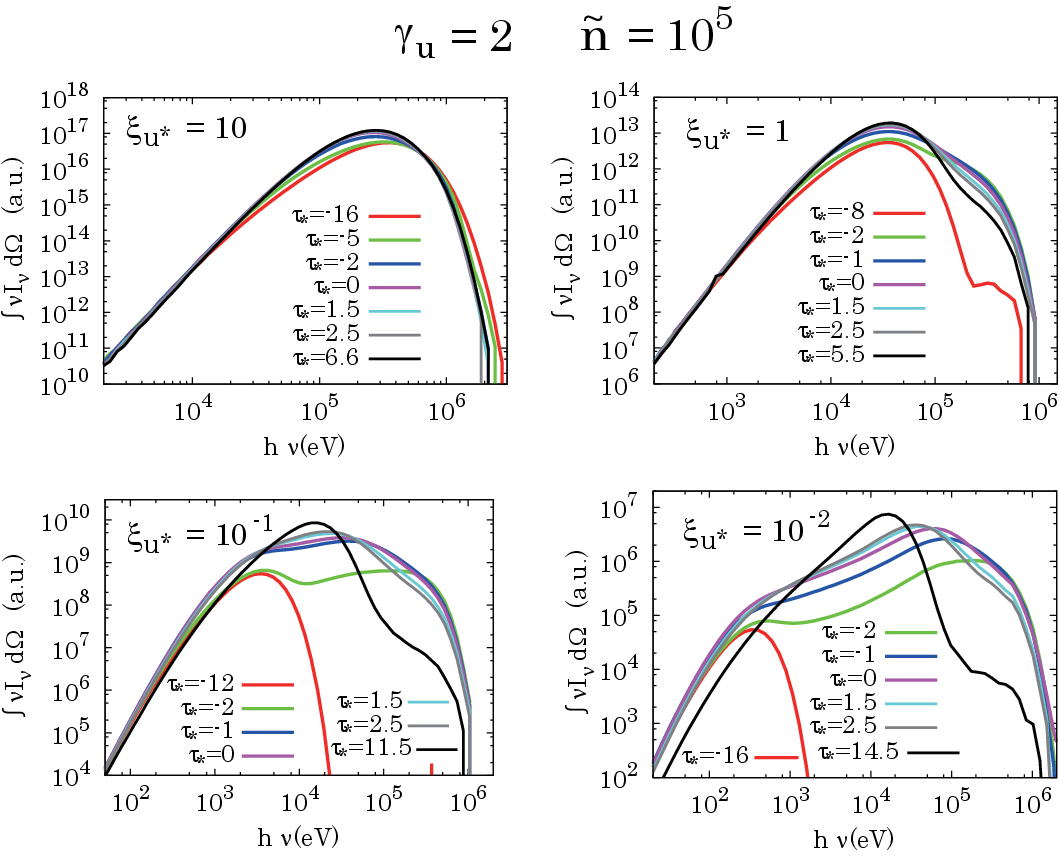}
\end{center}
\caption{Shock-frame, local angle integrated SEDs, 
 $\int \nu I_\nu(\tau*) d\Omega$, for models g2e1n5 ({\it top left}), g2e0n5 ({\it top right}), g2e-1n5 ({\it bottom left}) and g2e-2n5 ({\it bottom right}).
 The red and black lines show, respectively,  the spectra
 near the upstream and downstream boundaries of
 the simulation domain.  The value of  $\tau_*$ at the boundaries vary among the different models.
 The green, blue, magenta, cyan and gray lines display  
 spectra which were computed at locations $\tau_* = -2$, $ -1$, $ 0$, $1.5$ and $2.5$
 around the shock transition layer, as indicated.
 The scale on the vertical axis is given in arbitrary units. The absolute value can be specified once the
 number density of either baryons  or photons at far upstream ($n_u$ or $n_{\gamma, u}$) is specified.
 Note that the range of the horizontal axis is identical in all cases.
}
\label{Inug2n5dist}
\end{figure*}

\begin{figure}
\begin{center}
\includegraphics[width=8cm,keepaspectratio]{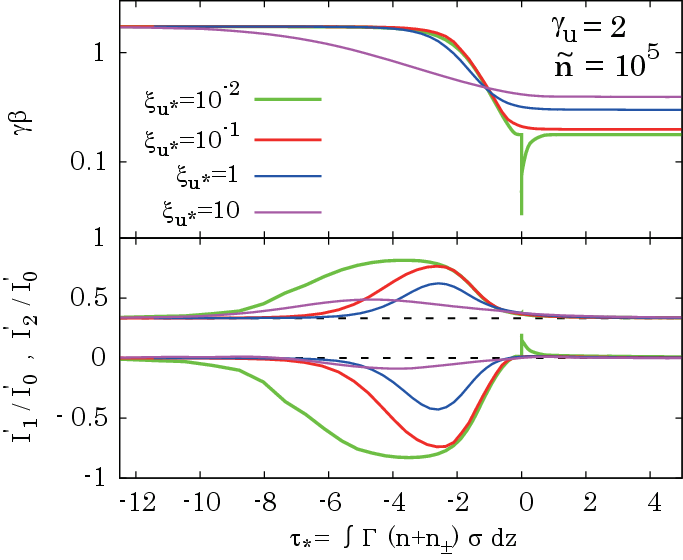}
\end{center}
\caption{Dependence of the 4-velocity profile ({\it top}) and 
 the normalized comoving  1st and 2nd moments of the radiation intensity, $I_{1}^{'}/I_{0}^{'}$ and   $I_{2}^{'}/I_{0}^{'}$ ({\it bottom}),
 on the far upstream photo-to-baryon inertia ratio $\xi_{u *}$,
  for  $\gamma_u = 2$ and $\tilde{n}=10^5$.
 The green, red, blue and magenta lines correspond to models g2e-2n5, g2e-1n5, g2e0n5 and g2e1n5, respectively.
 For a given pair of lines in each model in the bottom panel, the upper one corresponds to the second moment $I_{2}^{'}/I_{0}^{'}$, and the lower one to
 the first moment $I_{1}^{'}/I_{0}^{'}$.
 The two dashed lines in the bottom panel mark the values of the radiation moments of an isotropic radiation field ($I_{1}^{'}/I_{0}^{'}=0$ and $I_{2}^{'}/I_{0}^{'}=1/3$). 
}
\label{g2n5I}
\end{figure}

 As for  the fiducial models ($\gamma_u = 2$ and $\tilde{n} = 10^5$), 
 we  compute the cases for $\xi_{u*}=10, 1, 0.1$, and $10^{-2}$.
 The obtained shock structures
  are  displayed  in Fig. \ref{g2n5dist}.
 The horizontal axis in all plots shows the 
 angle averaged, pair loaded optical depth for Thomson scattering, as measured in the shock frame:
\begin{eqnarray}
\tau_{*} = \int \gamma (n + n_{\pm}) \sigma_T dz, 
\end{eqnarray} 
 where  $dz$ denotes the distance element along the flow direction.
It is measured from the subshock, when present, where $\tau_{*} = 0$,
and from the location where the bulk velocity has first reached the 
 far downstream value,  $\beta \simeq \beta_d$, when the subshock is absent.
 As a function of $\tau_{*}$, the vertical axis 
  shows the  4-velocity, $\gamma \beta$, 
 temperature, $T$,
 and the pair-to-baryon density ratio, $n_{\pm}/n$.
 Together with the plasma temperature, we also display the quantity 
\begin{eqnarray}
 T_{\gamma, eff} = \frac{I_0^{'}}{3 {\cal I}_0^{'}}
\end{eqnarray}  
 for reference, where $I_0$ and ${\cal I}_0$ are, respectively, the 0th moments of the intensity 
 $I_{\nu}$ and the photon flux density $I_{\nu}/h\nu$.
 Henceforth,  quantities with and without the superscript prime are measured
in the comoving frame  and shock frame, respectively.
 The $n$th moments of the intensity and photon flux density are defined as follows:
\begin{eqnarray}
I_{n} = 2 \pi \int \int I_{\nu} {\rm cos}^{n}\theta ~d\nu d\Omega, 
\end{eqnarray}  
\begin{eqnarray}
{\cal I}_{n} = 2 \pi \int \int \frac{I_{\nu}}{h \nu} {\rm cos}^{n} \theta ~d\nu d\Omega, ~~~(n=0,1,2),
\end{eqnarray}  
where $\theta$ is the angle between the flow velocity and the  photon direction. 
 Note that $T_{\gamma, eff}$ can be regarded as the actual temperature 
 of the radiation when the distribution is Wien, $I_{\nu}\propto \nu^3 {\rm exp}[h\nu/(kT)]$, or Planck.
 Henceforth, we refer to $T_{\gamma,  eff}$ as the effective radiation temperature.
 The angle integrated spectral energy distribution (SED),
 $\int \nu I_\nu d\Omega$,  computed in the shock frame at a given location, 
is exhibited in Fig.  \ref{Inug2n5dist} for each model at different locations. 
 In Fig. \ref{g2n5I} we show  a comparison of the 4-velocity profiles of the different fiducial models, 
 together  with the comoving 1st and 2nd moments of the radiation intensity 
 normalized by the 0th moment, $I_{1}^{'}/I_{0}^{'}$ and  $I_{2}^{'}/I_{0}^{'}$.
When the radiation field is isotropic in the comoving frame, we have $I_{2}^{'}/I_{0}^{'} =1/3$ and $I_1^{'}=0$, 
 while completely beamed radiation leads to
 $I_{2}^{'}/I_{0}^{'} =1$ and $I_1^{'}/I_{0}^{'}=-1$ or $1$.

 Although there is some difference between the models,
 the deceleration of the shock occurs over
 an optical depth $\tau*$ of a few in all cases. 
This stems from the fact that in the relativistic case the plasma crossing time of the shock is nearly equal 
 to the light crossing time. 
The relativistic motion of the plasma is also the sole reason why inside the shock transition layer 
the radiation appears highly anisotropic  in the rest frame of the fluid, as seen in Fig. \ref{g2n5I}.
 As mentioned in Section \ref{Intro},  this is in marked difference to 
 non-relativistic shocks in which the diffusion length is much longer, 
 and the radiation is nearly isotropic.
 During the deceleration, Compton scattering also
 heats up the plasma to higher temperatures. 
 Regarding the radiation spectra,
 while a Wien distribution at the local temperature is established
 at the far upstream and downstream  regions,
 non-thermal distributions originating from bulk Comptonization
are produced near the shock transition layer. 
%
 Apart from these general trends, the details of shock dissipation 
 vary considerably with $\xi_{u*}$.

 In model g2e1n5, 
 the inertia of the flow at far upstream 
 is largely dominated by the radiation  ($\xi_{u*}=10$).
 In this case,
 strong anisotropy cannot develop within the shock,
 since a small departure from  isotropy is
 sufficient to give significant impact on the bulk flow of the plasma.
 As a result, the velocity profile is 
 relatively smooth, reflecting gradual deceleration
 compared with the cases of lower $\xi_{u*}$. 
 As shown in \citet{B17},
 in the extreme limit of $\xi_{u*}=\infty$, the radiation field
 must satisfy the force-free condition $I_1^{'}=0$.
 Here (model g2e1n5) the plasma has a finite contribution to the inertia
 ($\sim 10 \%$ of the total), therefore a small but finite anisotropy is present.

 In this model, the temperature of the plasma coincides with
 the effective temperature of the radiation, $T_{\gamma, eff}$, at any position.
 This is due to the fact that  the photon distribution is close to Wien, so that
 Compton equilibrium (see Section \ref{Tcal} for details) is
 established throughout the shock.
 The spectra of photons do not largely depart from the Wien distribution because
 bulk Comptonization,  which mediates the shock, need not be significant.
 The resulting temperature shows gradual increase from $k T \sim 30~{\rm keV}$ to $\sim 60~{\rm keV}$ across the deceleration zone (see upper left panel of Fig. \ref{g2n5dist}).
%
 Although the change in the temperature is only a  
 factor of $\sim 2$ across the shock, significant increase  is found in the pair number density.
 This is because the pair production 
 rate by photons in a Wien distribution is a sensitive function of
 temperature in this  range ($k T \sim 30-60\,  {\rm keV})$,
 since only the  high energy population around the exponential cutoff 
 exceeds the threshold energy for pair creation.
 As a result,  the pair loading, $n_{\pm}/n$, increases
 by three orders of magnitude across the transition layer.
 At the far upstream and downstream regions, pair production and 
 annihilation are in balance and  the number density of pairs
 can be well approximated by that of Wien equilibrium at
 non-relativistic temperatures (see Section \ref{WIENeq}).
 From Equation (\ref{nonrelaWIEN}) ($n_{\pm}/n \sim \tilde{n} \Theta^{-3/2}{\rm exp}(-\Theta^{-1})$), we obtain
  $n_{\pm} \sim 0.35$ ($n_{\pm} \sim 6.2 \times 10^{2}$) 
 for a far upstream (downstream) temperature of $kT \sim 30~{\rm keV}$ ($\sim 60{\rm keV}$).
 Indeed, this is in good agreement with our simulation.

 As the value of $\xi_{u*}$ decreases, the velocity gradient,
 $d \gamma \beta / d\tau_{*}$,  steepens.
 This is mainly due to the fact that for a given density ratio $\tilde{n}$, the average energy of photons 
 decreases with  decreasing $\xi_{u*}$, since the upstream temperature satisfies $T_{u} \propto \xi_{u *}$.
 As a result, Klein-Nishina effects are diminished, and
 the average mean free path of photons  is reduced, ultimately approaching the Thomson limit. 
%
 A steeper velocity gradient is also required for lower  values of $\xi_{u *}$ in order
 to increase the efficiency at which the bulk kinetic energy is extracted by bulk Comptonization. 
 As shown in Fig. \ref{Inug2n5dist},
 when the photon-to-baryon inertia ratio is reduced to the value
 $\xi_{u *} = 10^{-2}$ (model g2e-2n5),
 a smooth velocity profile is no longer
 sufficient to achieve energy-momentum conservation to the required accuracy at every grid point, and
 our calculations imply the formation of a subshock in the system.
 It is noted, however, that the subshock is quite weak, in the sense that it carries only a small 
 fraction (a few percents at most)  of the entire shock energy, and so do not play an important role
 in the dissipation process. 
 Therefore, its impact on the radiation properties is also negligible.
 Therefore, in what follows we mainly focus on
 the global properties of the shock, that are not affected by the subshock.
 The details  of the subshock structure will be given later on, in Section \ref{weaksub}.  

The bulk Comptonization in the deceleration zone becomes significant
 as $\xi_{u *}$ decreases, and 
 results in the emergence of a non-thermal  spectrum.
 As shown in  Fig. \ref{Inug2n5dist}, 
 the spectral slope is harder for smaller values of $\xi_{u *}$.
 Concomitant with the hardening of the spectrum, 
 the departure from isotropy (as seen in the comoving frame) 
 that develops inside the shock 
 becomes more prominent (bottom panel of Fig. \ref{g2n5I}).

 The maximum energy attainable through bulk Comptonization 
 is limited by
 the  kinetic energy of the electrons/positrons to about
 $(\gamma_{u} - 1) m_e c^2 \sim 500~{\rm keV}$. 
 When the pair content is small, this corresponds roughly to the
 cutoff energy of the non-thermal photons at high energies  (e.g., models g2e0n5 and g2e-1n5).
 On the other hand, when gamma ray production via pair annihilation is important, 
 the conversion of rest mass energy leads to a moderate 
 increase in  the cutoff energy, roughly to
 $\gamma_{u} m_e c^2 \sim 1 ~{\rm MeV}$ (e.g., model g2e-2n5).
 Also note that, although the temperature is non-relativistic,
 thermal motions slightly shift the energy to higher values
 and  produce a broadening of the
 spectrum at the highest energies (Fig. \ref{Inug2n5dist}).

 Since Compton equilibrium is achieved  throughout the shock
 (except for the immediate post subshock region), as explained in Section \ref{Tcal}, and 
 higher energy photons  can exchange 
 their  energy more efficiently via scattering, 
 the presence of non-thermal photons will result in an abrupt 
 heating of the 
 plasma up to a temperature well in excess of $T_{\gamma, eff}$.
%
 Therefore, while no departure is found for model g2e1n5 ($T\sim T_{\gamma, eff}$),
 the deviation of the plasma temperature
 from  $T_{\gamma, eff}$ becomes more substantial
 as  $\xi_{u *}$  decreases (see Fig. \ref{g2n5dist}).
 This implies the presence of a prominent plasma heating precursor at the onset of
 the shock transition layer for relatively low values of $\xi_{u *}$.

 The pair density profile also changes significantly with $\xi_{u*}$.
 While there is a significant amount of pairs in model g2e1n5,
 they are negligible in models g2e0n5 and g2e-1n5 ($n_{\pm}/n \ll 10^{-10}$).
 This is a direct consequence of the
 lower peak energy (approximately $3 k T_{\gamma, eff}$), that gives rise to an 
 exponential suppression of the number of photons above 
  the pair creation threshold.
 On the other hand, while $T_{\gamma, eff}$ is still low, the production 
 of a prominent non-thermal component leads to enhanced 
 pair production in model g2e-2n5.
 The pairs only appear in the vicinity of the transition layer, since 
 the pair production opacity contributed by the bulk Comptonized photons peaks there.

 It should be noted that the existence of pairs can change
 the spatial width of  the shock considerably once their density exceeds the baryon density ($n_{\pm}/n \gtrsim 1$)
 and begins to govern the scattering opacity inside the shock.
 For example, the physical length scale per optical depth $dz/d\tau_{*}$
 at far upstream is longer than that of the far
 down stream by roughly 3 orders of magnitude.
 Therefore, one should bear in mind that, while the shock width 
 in terms of $d\tau_{*}$ is similar among the models,
 it could largely differ when measured in terms of the physical length scale $dz$, even when the
 same far upstream density $n_u$ is invoked.

 \subsubsection{Emergence of a weak subshock}
 \label{weaksub}

\begin{figure}
\begin{center}
\includegraphics[width=7cm,keepaspectratio]{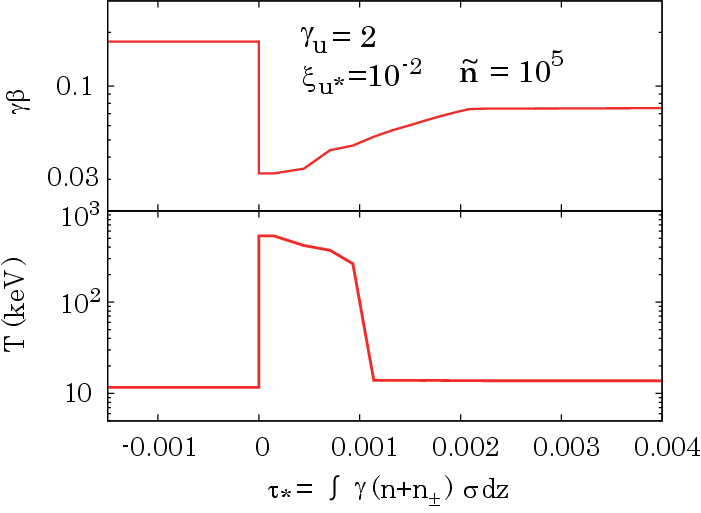}
\end{center}
\caption{Enlarged view of the 4-velocity and temperature profiles around the weak subshock,
for model g2e-2n5.
}
\label{g2e-2n5sub}
\end{figure}

 As mentioned earlier, emergence of a
 weak subshock seems necessary in model g2e-2n5.
 Although its contribution to the overall dissipation is quite
 small, its existence is required to achieve steady flow
 solutions (see  Section \ref{Numsetup} for 
 details).
 As described in Section \ref{Numsetup},
 we treat the subshock as a discontinuity
 in the flow parameters
 that satisfy the Rankin-Hugoniot condition for a plasma isolated from the radiation.
 A notable feature of the subshock is a sharp
 spike followed by a dip in the velocity and temperature profiles. 
 The drop in the velocity to a value smaller than the
 far downstream velocity is an inevitable consequence of the 
 plasma sound speed, $c_s \approx [5 P_{pl}/3\rho_{pl}]^{1/2}$,
 being small
 ($c_s/c \sim 0.09$ for $kT\sim 500~{\rm keV}$ and $n{\pm}/n \sim 10$).
 The rise of the temperature just behind the subshock, up to 
 $k T_{d,sub} \sim 500~{\rm keV}$, 
 is caused by the  self-generated heat of the plasma within the subshock.
 Since the photons cannot interact with particles over the 
 plasma scale, the post shock temperature is well above that obtained in 
 Compton equilibrium.
 Consequently, following  shock heating, the pairs exposed to
 the intense radiation field rapidly cool via Compton scattering
 until the temperature reaches the equilibrium value 
 (roughly equals to that ahead of the subshock).
 As a result, a structure that resembles an isothermal shock is formed
 (see a magnified view in Fig. \ref{g2e-2n5sub}). 
Within the cooling layer ($\tau_{*} \lesssim 0.001$), the bulk plasma rapidly
 accelerates,  predominantly by its pressure gradient force. 
 Above the cooling layer, the acceleration continues more gradually, 
 mainly due to  the radiation force, up to the distance  where
 it reaches the far downstream velocity (at $\tau_{*} \sim 0.6$)

 A crude evaluation of the thickness of the cooling layer, $d\tau_{*,cool}$,
 can be derived as follows:
 The number of scatterings per unit time for a
 single electron/positron  is given by $\sim n_{\gamma}c \sigma_T$
 in the comoving frame.
 Hence, given the energy loss per scattering, 
 $\sim 4 <h\nu> k T_{d,sub} / m_e c^2 $,
 and the downstream thermal energy per electron/positron, $3 k T_{d,sub}$,
 the cooling time is derived as
 $t_{cool} \sim 3/4 (<h\nu>/ m_e c^2)^{-1}  (n_{\gamma}c \sigma_T)^{-1}$,
 where $<h\nu>$ and $k T_{d,sub}$ denote the
 average photon energy and the temperature at the immediate downstream of the subshock.
 In terms of the effective temperature,
 the average photon energy at the subshock can be expressed as 
 $<h\nu> \sim 3 k T_{\gamma, eff}$.
 While the photon number density is approximately 
 $n_{\gamma} \sim \tilde{n}n$ over most of the RRMS layer,
  it is given by
 $n_{\gamma} \sim \tilde{n}n \beta_{u,sub}/\beta_{d,sub}$ at 
 the immediate downstream of the subshock, owing to the  sudden compression of the plasma there, 
 where $\beta_{u,sub}$ ($\beta_{d,sub}$) is the velocity at the immediate upstream (downstream) of the subshock.
%
 Taking into account the above factors, the cooling layer thickness can be expressed as
\begin{eqnarray}
\label{coollength}
 d\tau_{*,cool}  &\sim &
              \beta_{d,sub} c (n+n_{\pm}) \sigma_T t_{cool}  \\  
                 &\sim &
             1.5 \times 10^{-3}
                  \left(\frac{kT_{\gamma,eff}}{2~{\rm keV}}\right)^{-1} 
 		  \left(\frac{\tilde{n}}{10^5} \right)^{-1}
                   \left(\frac{ \frac{n+n_{\pm}}{n}}{10}\right) \nonumber \\ 
                 && ~~~~~~~~~~ \times
		  \left(\frac{\frac{\beta_{u,sub}}{\beta_{d,sub}}}{8}\right) 
		  \left(\frac{\beta_{d,sub}}{0.03}\right) . \nonumber
\end{eqnarray}
%
Pair creation and annihilation were  ignored in the above derivation, as they are negligible over 
 the cooling layer
 given its small thickness relative to the entire RRMS transition layer (see bottom right panel of Fig. \ref{g2n5dist}).

 We can confirm from  Fig. \ref{g2e-2n5sub} 
 (as well as from Figs. \ref{g2n3dist}, \ref{g4e-2n5sub} and \ref{g10n5sub}
 for the other models with subshocks)
 that this rough estimation is in agreement
 with our numerical results within a factor of a few.
 Note that there are several factors
that were ignored in our crude estimation of the
cooling layer thickness, and which can lead to some
differences between the analytic and numerical results.
%
 For example, we have neglected the effect of adiabatic cooling as well as 
 the effect of broad radiation spectrum.
 Moreover, in evaluating the cooling rate, we have used an expression which 
 is only valid in the non-relativistic limit, $kT_{u,sub},kT_{d.sub} \ll m_e c^2$, while the temperature is typically mildly relativistic.
In view of these simplifications, we find the mild disagreement 
between the numerical result and the analytic result derived above reasonable.

 It is worth noting that,
 while this weak subshock strongly affects the properties of the plasma in its vicinity,
 it has almost no  influence on the radiation.
 This is simply because the thermal energy generated by the subshock,
 $3 (n + n_{\pm}) k T_{d,sub}$,
 is negligible compared with  that contained in the radiation,
 $3 n_{\rm \gamma} k T_{\gamma, eff}$.
 Therefore, the weak subshock does not affect
 the overall energetics of the system nor the radiation properties.
 This is also true for all the other cases in which subshocks were found, and for which 
 the photon-to-baryon number ratio is sufficiently above the critical value
 $\tilde{n}_{crt}$ given in Equation (\ref{ncrit})
 (see Section \ref{sec:photon_rich}).

 While our numerical simulations predict  their existence,
 we could not identify the physical origin of 
 the ``weak'' subshocks  that we found in the regime $\tilde{n} \gtrsim n_{crt}$
 (models g2e-2n5, g4e-2n5, g10e-1n5 and g10e-2n5),
 unlike the case of a photon starved shock, $\tilde{n} < n_{crt}$ (models g2e-1n3 and g2e-2n3),
  where formation of a  ``strong'' subshock is dictated by inefficient energy extraction thorough 
  Compton scattering, as will be discuss in greater detail  in Section \ref{ndepend} below.
Though non trivial, this presumably indicates that
 no steady, continuous flow  solutions exist in a certain regime of the parameter space.
  As seen in Fig. \ref{g2n5I}, the flow velocity gradient tends
 to steepen as the value of $\xi_{u *}$ is reduced. 
 Our result suggests that there is a threshold value of $\xi_{u *}$ 
 below which the continuous steepening of the
 velocity profile ultimately turns into a weak subshock at the
 edge of the RRMS transition layer.
%
 It is worth mentioning that \citet{BKAW10} also found 
 a weak subshock in their simulations of photon starved RRMS (in which photon generation is included). 
%
%
 It should be stressed, however, that these weak subshocks are merely 
 small disturbances  in the global shock structure, and their physics 
 is not important in evaluating the overall dynamics of the bulk flow
 as well as the radiation properties.

\subsection{Dependence on $\tilde{n}$}
\label{ndepend}

\begin{figure*}
\begin{center}
\includegraphics[width=15cm,keepaspectratio]{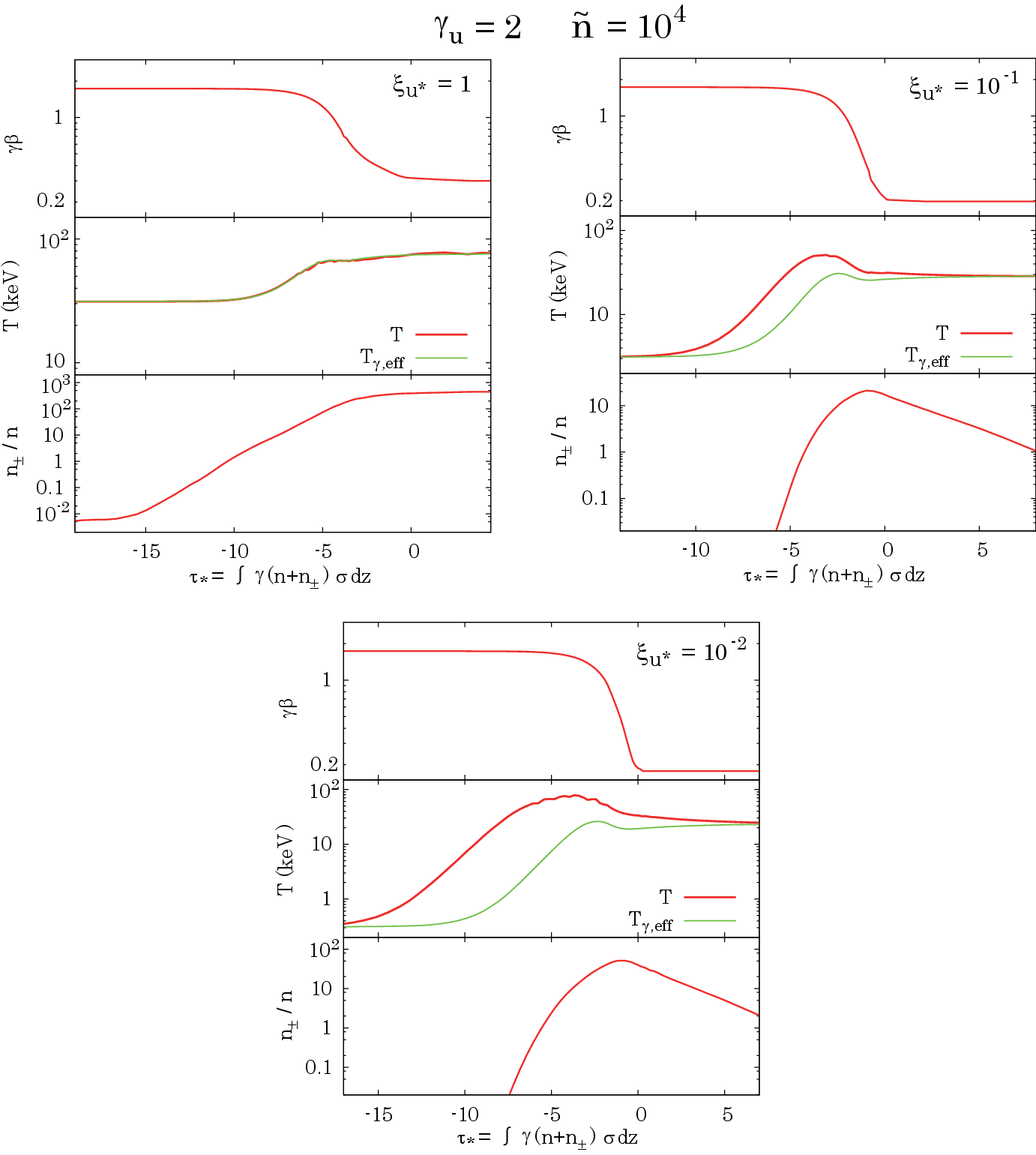}
\end{center}
\caption{Same as Fig. \ref{g2n5dist}, but for g2e0n4 ({\it top left}), g2e-1n4 ({\it top right})  and g2e-2n4 ({\it bottom}). 
}
\label{g2n4dist}
\end{figure*}

\begin{figure*}
\begin{center}
\includegraphics[width=14cm,keepaspectratio]{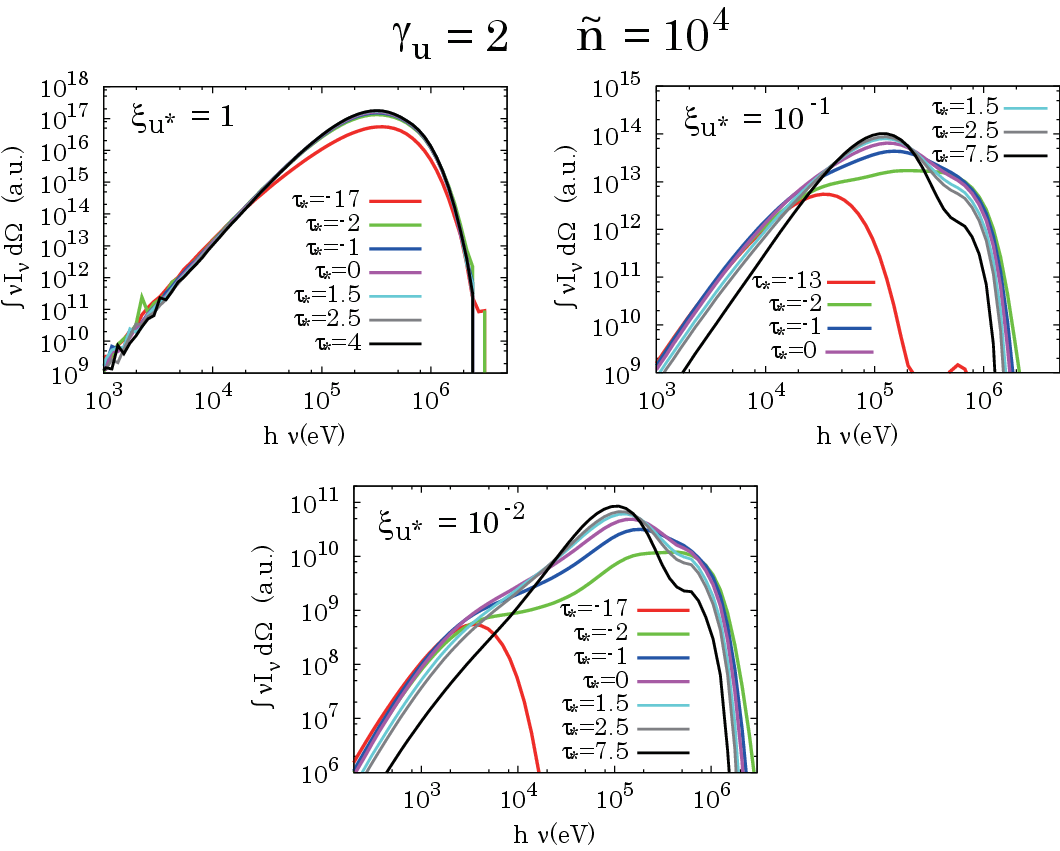}
\end{center}
\caption{Same as Fig. \ref{Inug2n5dist}, but for g2e0n4 ({\it top left}), g2e-1n4 ({\it top right})  and g2e-2n4 ({\it bottom}).
}
\label{Inug2n4dist}
\end{figure*}

To explore the dependence of the shock properties 
on the  photon-to-baryon number ratio, we performed several
calculations that invoke smaller values of $\tilde{n}$ ($10^4$ and $10^3$)
than that used in the fiducial models, but the
same values of $\gamma_u$ and $\xi_{u*}$. 
In the models with $\tilde{n}=10^4$, three cases
with different values of photon-to-baryon inertia ratio 
($\xi_{u *} = 1$, $0.1$ and $0.01$) are considered
(g2e0n4, g2e-1n4 and g2e-2n4).
Their overall structures and spectra are summarized in 
Figs. \ref{g2n4dist} and  \ref{Inug2n4dist}.
%
As seen, the general trends are quite similar to those of  the fiducial models;
the decrease in $\xi_{u *}$ results in a steepening of
their velocity gradient $d\gamma \beta/d\tau_{*}$ 
and in the enhancement of the non-thermal spectrum.


Apart from the similarities, there are also interesting differences from the 
fiducial models ($\tilde{n} = 10^5$).
For a fixed value of $\xi_{u *}$, lower $\tilde{n}$ results in a 
higher temperature ($T \propto \tilde{n}^{-1}$), since the same amount of energy is shared by a smaller number of particles (photons, protons and pairs).
Hence, the overall temperature and average photon energy
are roughly $ 10$ times higher in these models.
This  shifts the average mean free path of photons to larger values
owning to the increase in the population of photons that are scattered in
the Klein-Nishina regime.
As a result, the  deceleration lengths 
 are found to be longer than those in
 the corresponding fiducial models
(g2e0n5, g2e-1n5, g2e-2n5).
%
The higher temperature and photon energy are probably the reason for 
 the absence of a weak subshock in model g2e-2n4,
 in difference from model g2e-2n5
 (that has the same  $\xi_{u *}$ value).
 We speculate that the smoother velocity profile in model g2e-2n4, that results
 from the larger penetration depth of the photons, enables the existence of steady solutions 
 with no subshock.
However, it is expected that a weak subshock will form also in these models for 
sufficiently low values of  $\xi_{u *}$ ($< 0.01$).

The larger temperature or, equivalently,  average photon energy,
also leads to enhanced pair production rate.
In particular, while the pair content is negligible
for $\xi_{u * }=1$ and  $\xi_{u * }=0.1$ in the fiducial models
(g2e0n5 and g2e-1n5), the models with $\tilde{n}=10^4$ and the same $\xi_{u * }$ values (g2e0n4 and g2e-1n4)
give rise to a significant amount of pairs.
Likewise, the pair density in model g2e-2n4 is higher by an order of magnitude 
than that  in the fiducial model g2e-2n5.

Comparing the structures,
the profiles in model g2e0n4 are similar to those
in model g2e1n5, rather than in 
 model g2e0n5 that has the same  $\xi_{u *}$ value.
Accordingly, as in  model g2e1n5,
the radiation and pairs are well approximated to be in
Wien equilibrium at far upstream and downstream,
while the in the transition layer they depart from the equilibrium due to a slight deviation from  the Wien distribution.
On the other hand, the shape of the spectrum in
 models g2e-1n4 and g2e-2n4 is similar to that of the counterpart fiducial models with same $\xi_{u *}$
 (g2e-1n5 and g2e-2n5),
 but its average energy is shifted toward higher energies, by a factor of $\sim 10$, while the cutoff energy remains unchanged
$\sim \gamma_u m_e c^2 \sim 1~{\rm MeV}$.
The higher photon energies in models g2e-1n4 and g2e-2n4 give rise
to a higher pair production rate than in the fiducial models.
Therefore, in all of these models, we find a non-negligible pair content.

\begin{figure*}
\begin{center}
\includegraphics[width=15cm,keepaspectratio]{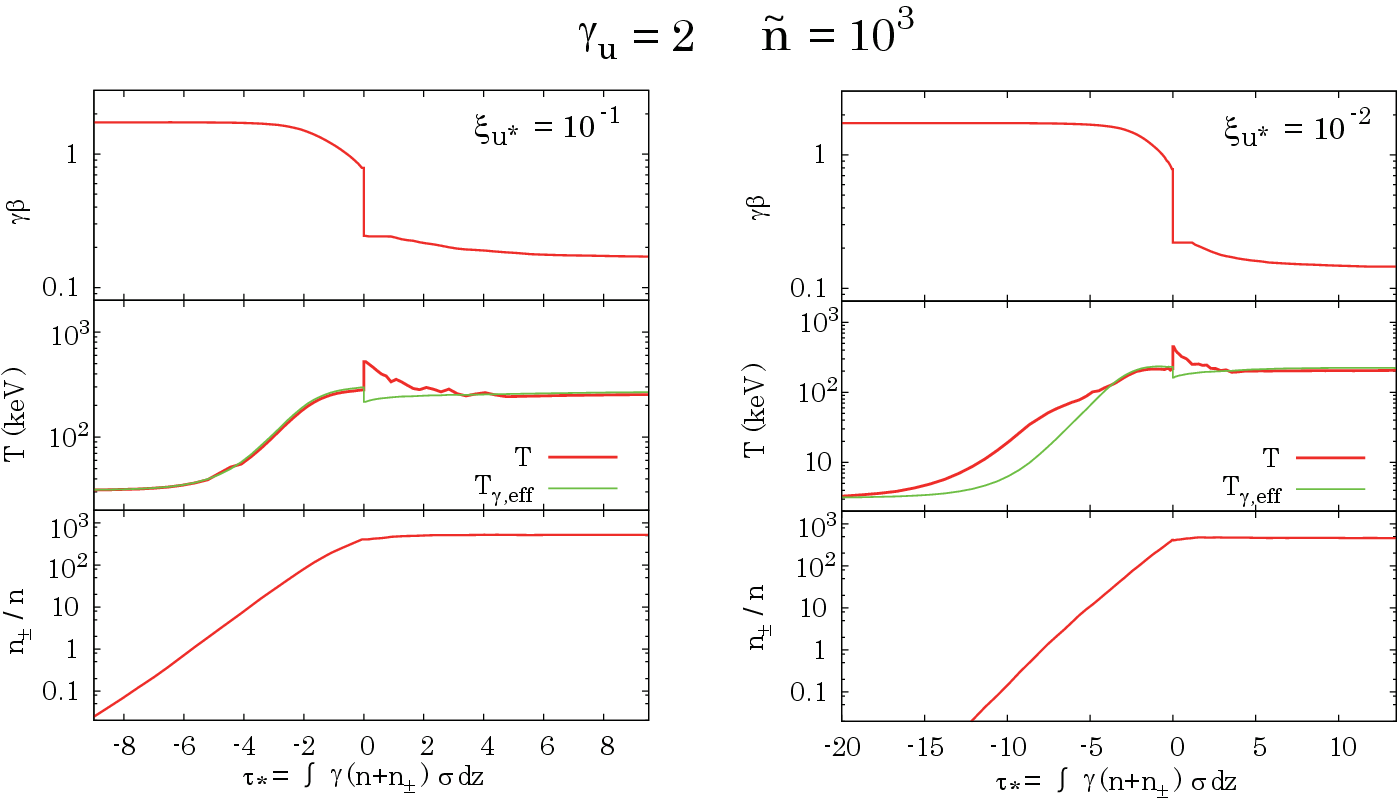}
\end{center}
\caption{Same as Fig. \ref{g2n5dist}, but for g2e-1n3 ({\it left})  and g2e-2n3 ({\it right}). 
}
\label{g2n3dist}
\end{figure*}

\begin{figure*}
\begin{center}
\includegraphics[width=14cm,keepaspectratio]{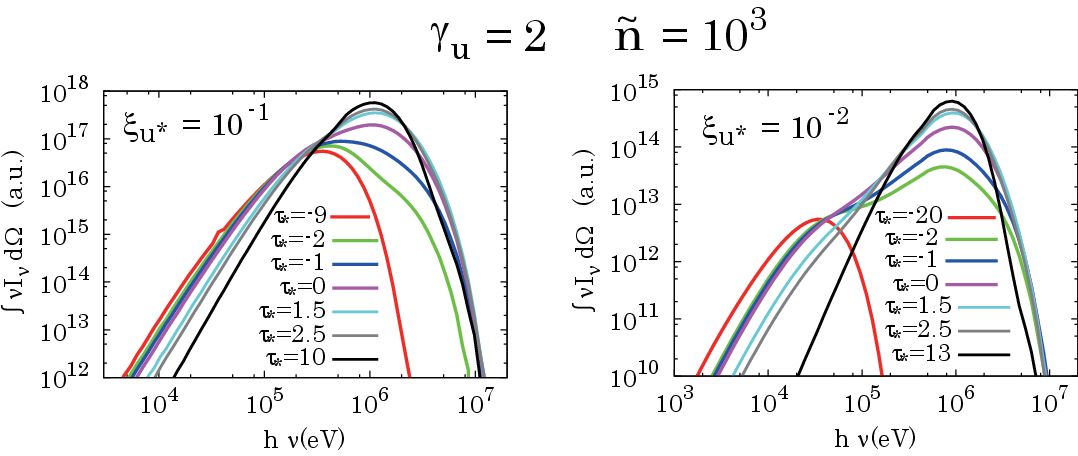}
\end{center}
\caption{Same as Fig. \ref{Inug2n5dist}, but for g2e-1n3 ({\it left})  and g2e-2n3 ({\it right}).
}
\label{Inug2n3dist}
\end{figure*}

The properties of the shocks drastically change in the models with
$\tilde{n}=10^3$.
In the present study,
two cases with the values $\xi_{u *} = 0.1$ and $0.01$ are computed (g2e-1n3 and g2e-2n3).
Their overall structures and spectra are exhibited in 
Figs. \ref{g2n3dist} and  \ref{Inug2n3dist}, respectively.
The notable difference from the models with higher $\tilde{n}$ 
 is the  formation of a ``strong'' subshock.
 Unlike  the ``weak'' subshocks found in some of the other models
 (see Section \ref{weaksub} for details),
 the physical origin of the strong subshocks is understood, and will be described 
 in detail  in Section \ref{Trans}.  

As Equation (\ref{shock_rich_temp}) predicts, the temperature downstream of the subshock in the models with $\tilde{n}=10^3$ approaches the pair equilibrium
value, $kT \sim 200~{\rm keV}$, as seen in Fig. \ref{g2n3dist}. 
The pair density increases rapidly inside the shock and approaches 
 the Wien equilibrium value, $n_{\pm} \approx n_{\gamma} K_2(\Theta^{-1})/ \Theta^2$, just downstream of the subshock. 
 At this temperature the pair density becomes comparable to the photon density, $n_\pm \sim n_{\gamma}$.  Since 
 in the absence of an internal photon source the number of quanta is conserved, we have $\tilde{n}=(n_\gamma+n_\pm)/n$ in the downstream region,
 which effectively reduces the number of photons that can extract energy, and strengthens the subshock further.
 One should keep in mind that while the subshock is relatively strong, it dissipates only about 30\% of the entire shock energy (in model g2e-2n3).
 Thus, a moderate increment in the photon density downstream (by no more than a factor of a few) will considerably 
 weaken, or completely eliminate, the subshock.  We anticipate
 this to happen once internal photon sources (in particular free-free emission by the hot pairs) are included.

\begin{figure}
\begin{center}
\includegraphics[width=8cm,keepaspectratio]{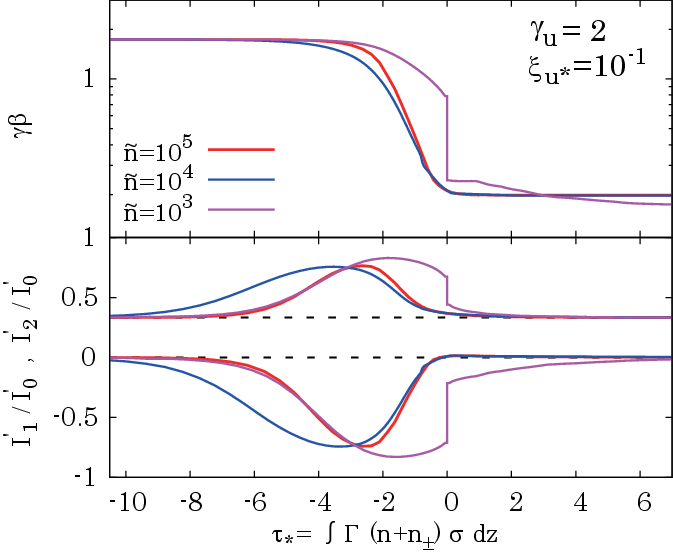}
\end{center}
\caption{Same as Fig. \ref{g2n5I}, but for  models g2e-1n5 (red), g2e-1n4 (blue) and g2e-1n3 (magenta).
}
\label{g2xi-1I}
\end{figure}

Shock solutions that correspond to the fiducial models with fixed $\xi_{u*}=0.1$ are compared,
for clarity, in Fig. \ref{g2xi-1I}.  The distinct properties of the marginally starved shock (g2e-1n3)
stand out.   The discontinuity in the profiles of the moments $I_{1}^{'}/I_{0}^{'}$ and  $I_{2}^{'}/I_{0}^{'}$
in the marginally starved shock  arises from the sudden change in the velocity 
of fluid elements (and, hence, in the frame in which these moments are computed) across the subshock.

\subsubsection{Transition to the photon starved regime}
\label{Trans}
Next, let us examine the transition from the photon rich to
the photon starved regime in some greater detail. 
In section \ref{sec:photon_rich} it has been argued that  once the photon-to-baryon number ratio far upstream becomes smaller than the critical 
value $\tilde{n}_{crt}$,  the advected photons cannot support the shock anymore, and the shock becomes photon starved.   In the absence 
of photon production processes one  expects that the strength of the subshock will dramatically increase as $\tilde{n}$ approaches $\tilde{n}_{crt}\simeq 10^3$.   This is 
the situation in models g2e-1n3 and g2e-2n3.    Fig. \ref{g2n3dist} exhibits results obtained for these models, verifying that the subshock is indeed substantially stronger than in the runs with $\tilde{n}>\tilde{n}_{crt}$.   As also seen, the downstream temperature reaches $200$ keV (except for the spike produced by the subshock), 
in accord with Equation (\ref{shock_rich_temp}),
leading to a vigorous pair creation in the shock transition layer.   The pair-photon plasma downstream quickly reaches equilibrium, with roughly equal densities, $n_\pm/n_\gamma\simeq1$.

From Equation (\ref{delta_n/n_gamma}) it is anticipated that under these conditions photon generation (not included in our simulations) will start dominating over photon 
advection, so that in reality the shock will be supported by photons produced inside and just behind the shock, and the subshock will disappear or remain insignificant.   
For higher upstream Lorentz factors, $\gamma_u \gg 1$, we expect that photon generation will dominate at somewhat higher $\tilde{n}$ values, roughly by a factor of $\gamma_u/2$, 
since even though the shock can be supported by the advected photons the temperature downstream exceeds the pair production threshold, at which e$^\pm$ pair  equilibrium is established.   The results of \citet{BKAW10} 
 confirm this.   We are currently in the process of modifying the code to include free-free and 
double Compton emissions.  Results of simulations of photon starved shocks will be presented in a future publication.

\subsection{Dependence on $\gamma_u$}

To investigate the dependence of the shock properties 
on the Lorentz factor, we have calculated
two sets of models with higher $\gamma_u$ ($4$ and $10$),
but with the
values of $\tilde{n}$ and $\xi_{u*}$ being
identical to those in the fiducial models.
In both cases, three calculations that invoke
different $\xi_{u*}$ values ($1$, $0.1$ and $0.01$) were performed, and are described next.

\begin{figure*}
\begin{center}
\includegraphics[width=15cm,keepaspectratio]{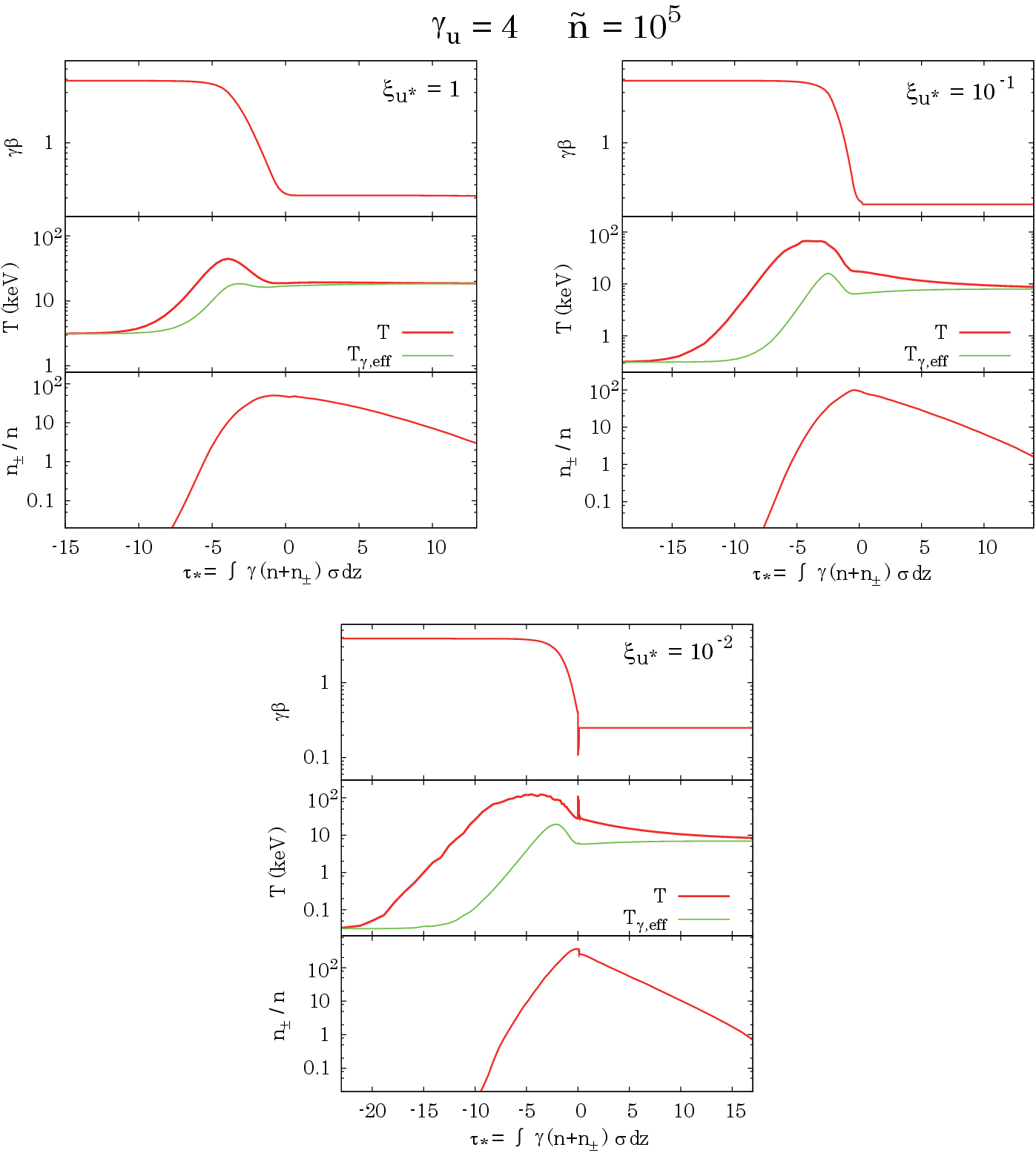}
\end{center}
\caption{Same as Fig. \ref{g2n5dist}, but for g4e0n5 ({\it top left}), g4e-1n5 ({\it top right})  and g4e-2n5 ({\it bottom}). 
}
\label{g4n5dist}
\end{figure*}

\begin{figure*}
\begin{center}
\includegraphics[width=14cm,keepaspectratio]{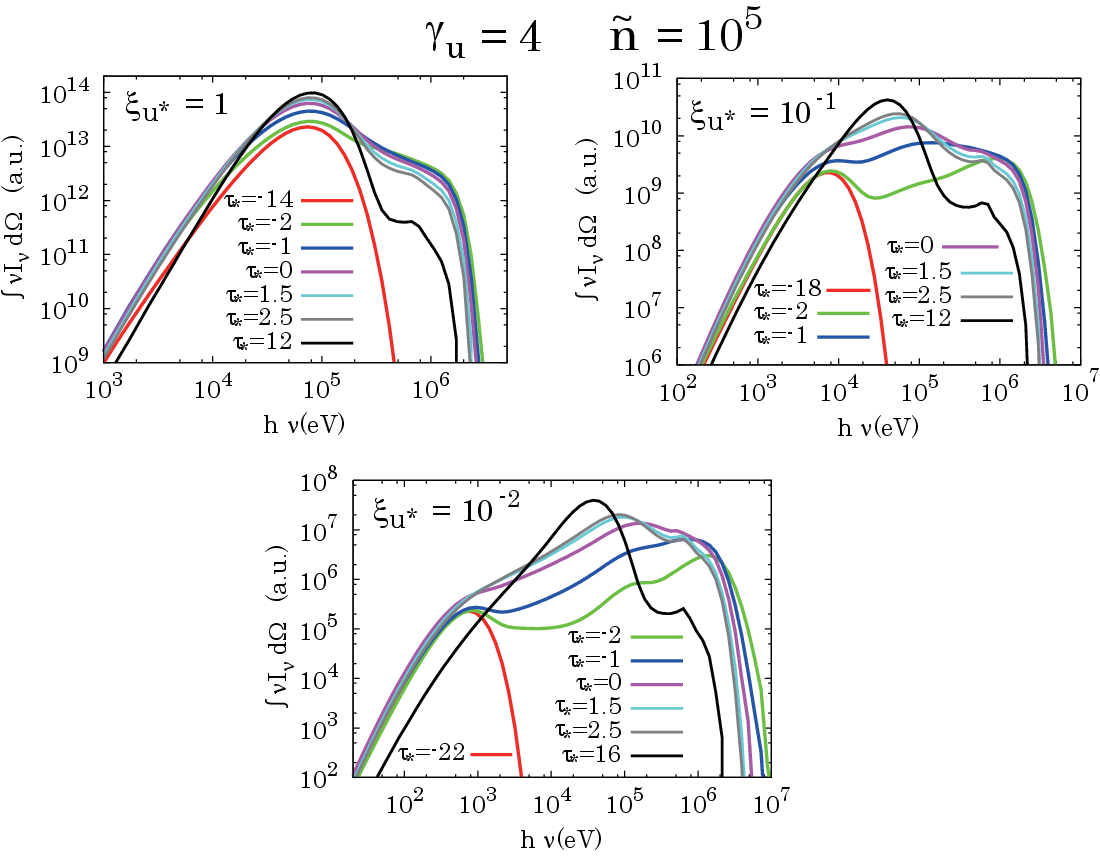}
\end{center}
\caption{Same as Fig. \ref{Inug2n5dist}, but for  g4e0n5 ({\it top left}), g4e-1n5 ({\it top right})  and g4e-2n5 ({\it bottom}). 
}
\label{Inug4n5dist}
\end{figure*}

\begin{figure}
\begin{center}
\includegraphics[width=7cm,keepaspectratio]{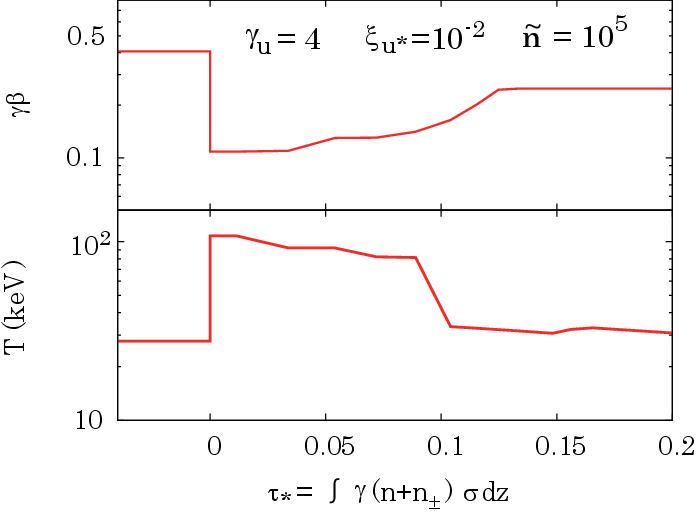}
\end{center}
\caption{Enlarged view of the 4-velocity and temperature profiles around the weak subshock region
for model g4e-2n5.
}
\label{g4e-2n5sub}
\end{figure}

\begin{figure*}
\begin{center}
\includegraphics[width=15cm,keepaspectratio]{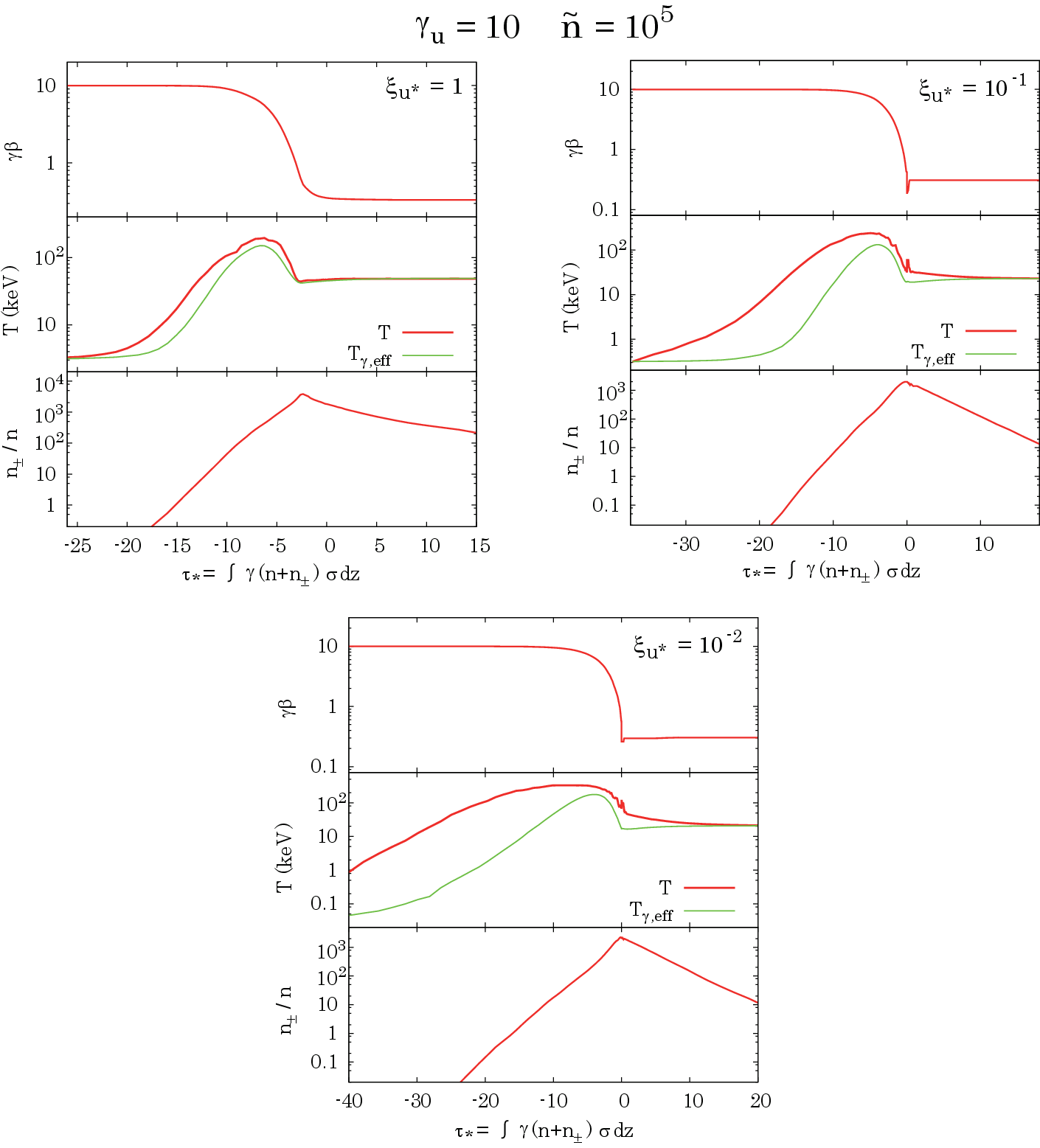}
\end{center}
\caption{Same as Fig. \ref{g2n5dist}, but for g10e0n5 ({\it top left}), g10e-1n5 ({\it top right})  and g10e-2n5 ({\it bottom}). 
}
\label{g10n5dist}
\end{figure*}

\begin{figure*}
\begin{center}
\includegraphics[width=14cm,keepaspectratio]{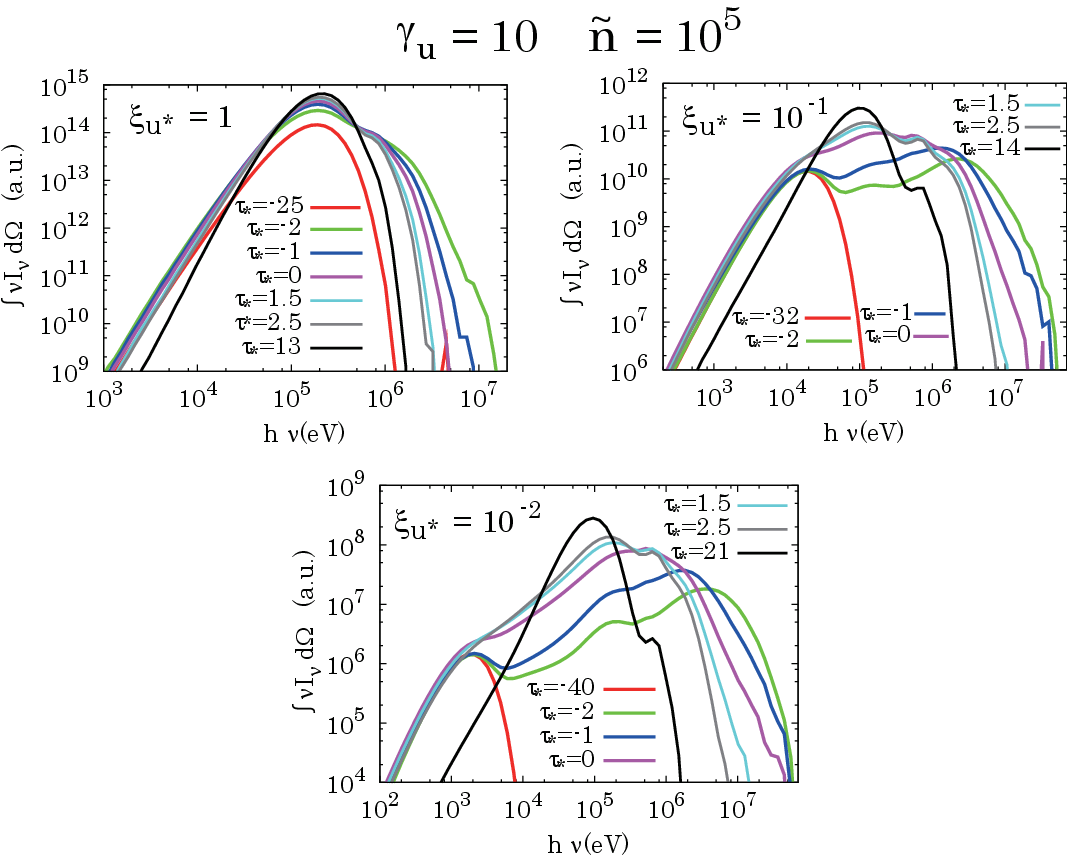}
\end{center}
\caption{Same as Fig. \ref{Inug2n5dist}, but for  g10e0n5 ({\it top left}), g10e-1n5 ({\it top right})  and g10e-2n5 ({\it bottom}). 
}
\label{Inug10n5dist}
\end{figure*}

\begin{figure*}
\begin{center}
\includegraphics[width=14cm,keepaspectratio]{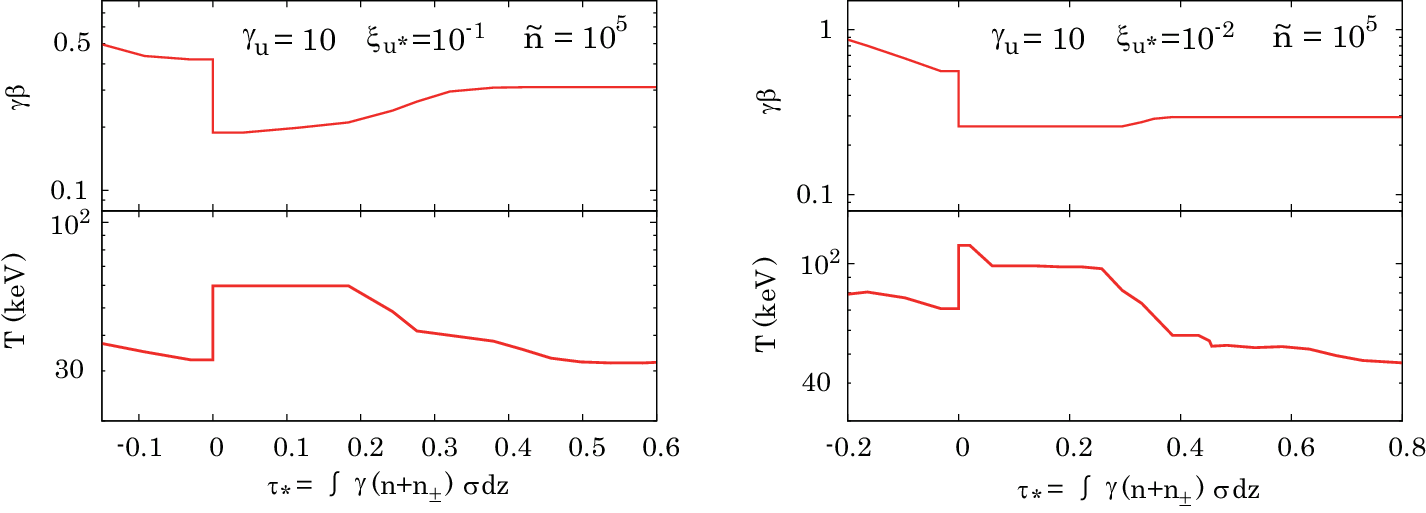}
\end{center}
\caption{Enlarged view of 4-velocity and temperature profile around the weak subshock region
for model g10e-1n5 and g10e-2n5.
}
\label{g10n5sub}
\end{figure*}

The structures and spectra obtained in the models with $\gamma_u = 4$ and $\gamma_u = 10$
are displayed in Figs. \ref{g4n5dist} - \ref{g10n5sub}.
Like in the fiducial models, also here
the velocity profile steepens as $\xi_{u*}$ is reduced.
The trends of the temperature profile are also similar to those in the fiducial models,
albeit with a higher downstream temperature, since it is roughly proportional
to the 4-velocity far upstream when $\xi_{u *} \lesssim 1$
 (see Equation (\ref{shock_rich_temp})).
%
%
At low values of $\xi_{u*}$ a weak subshock appears (see Figs. \ref{g4e-2n5sub} and  \ref{g10n5sub}  for a magnified view), as in the fiducial models.
The larger $\gamma_u$ the larger the value of $\xi_{u*}$ at which the subshock forms ($\xi_{u*}\le0.01$ for $\gamma_u=4$ and $\xi_{u*}\le 0.1$ for
$\gamma_u=10$).  The reason for this is unclear at present.  It might be related to the fact that the condition for starvation is proportional to $\gamma_{u}$ 
(see Equation (\ref{pht-starv-nd}) ).

The main effects caused by increasing the shock Lorentz factor
can be observed in the resulting spectra and pair populations, and 
can be summarized as follows: (i) The heating precursor broadens 
and the peak temperature increases as $\gamma_u$ increases, and likewise
the width of the shock transition layer.
(ii)  The pair content rises sharply as $\gamma_u$ increases, as is evident
from a comparison of Figs. \ref{g4n5dist} and \ref{g10n5dist}.
This is a direct consequence
of the fact that the number of bulk Comptonized photons that surpass the pair production threshold and,
hence, the pair production rate, are sensitive  functions of $\gamma_u$.
The large pair enrichment gives rise to a pronounced
signature of the $511~{\rm keV}$ pair annihilation line in the 
spectrum  (the small spectral bumps seen in Figs. \ref{Inug4n5dist} and \ref{Inug10n5dist}).
(iii) For fixed values of $\tilde{n}$ and $\xi_{u*}$ the high energy cutoff of the spectrum is 
roughly proportional to $\gamma_u$, as naively expected.

\begin{figure}
\begin{center}
\includegraphics[width=8cm,keepaspectratio]{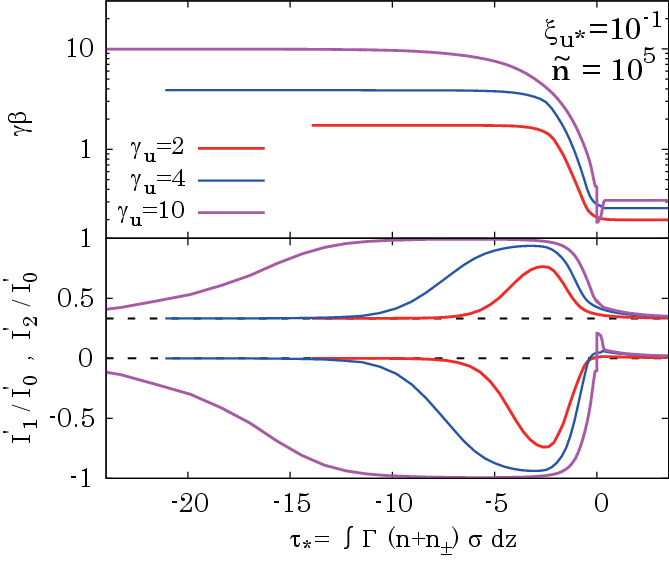}
\end{center}
\caption{Same as Fig. \ref{g2n5I}, but for model g2e-1n5 (red), g4e-1n5 (blue) and g10e-1n5 (magenta).
}
\label{xi-1n5I}
\end{figure}

To summarize the dependence of the shock structure on the bulk Lorentz factor,
we compare, in Fig. \ref{xi-1n5I}, the profiles of $\gamma \beta$,
 $I_{1}^{'}/I_{0}^{'}$ and  $I_{2}^{'}/I_{0}^{'}$
in the three models (g2e-1n5, g24-1n5 and g10e-1n5) that
 have different values for $\gamma_u$ 
 but same values  of  $\xi_{u*}$ ($=0.1$) and 
 $\tilde{n}$ ($=10^5$).
 As seen, the shock width slowly increases with increasing $\gamma_u$, in rough agreement with
 the analytic solution derived in Section \ref{sec:analyt}.  The level of anisotropy of the photon distribution and its extent
 also become larger as $\gamma_u$ is increased. 
 In the highest Lorentz factor case, the radiation intensity 
 achieves nearly complete beaming ($I_{2}^{'}/I_{0}^{'}=1, I_{1}^{'}/I_{0}^{'}=-1$).
 This reflects the rise in the population of high energy photons that penetrate against the flow to larger distances upstream.

\section{Applications}
\label{application}

So far, we have focused on the fundamental properties of RRMSs.
Here let us consider the applications  to GRBs.

Since RRMSs are expected to form in sub-photospheric regions,
they should have  substantial imprints on the resulting emissions   \citep{BML11, LE12, KL14}.
As shown in the previous section, when the energy density of the radiation at far upstream  is much larger than that of the 
rest mass energy of the plasma, viz., $\xi_{u *} \gg 1$,
thermal spectra with roughly the same peak energy and flux
are found at any location in the shock  (see top left panel of Fig. \ref{Inug2n5dist}).
This implies that observed spectra produced by a sub-photospheric shock (even if strong)  should be 
nearly thermal when the photosphere is located far below the saturation radius.

\begin{figure*}
\begin{center}
\includegraphics[width=15cm,keepaspectratio]{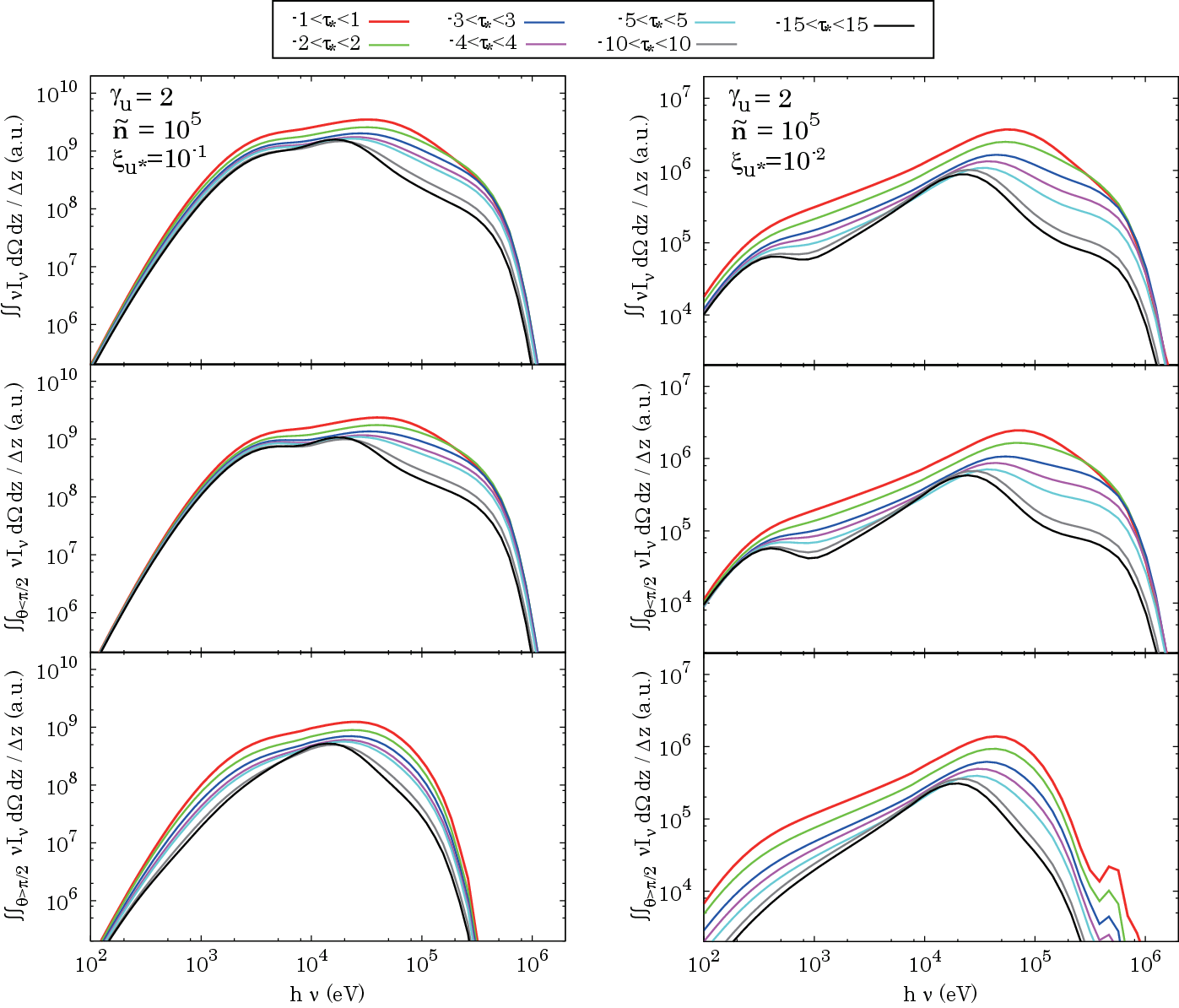}
\end{center}
\caption{Shock-frame, angle integrated spectra, $\int \nu I_\nu d\Omega dz $,
 averaged over physical distance within a given interval $\Delta \tau*$,
  for models g2e-1n5 (left) and g2e-2n5 (right).
The upper panels show spectra that were integrated over all directions. The
 middle panels display spectra that were integrated only over photons that propagate
 toward the downstream  ($\theta \leq \pi / 2$), and the bottom panels over photons that propagate toward the upstream ($\theta > \pi / 2$).
The different colors correspond to the special domain over which each spectrum was computed, with red, green, blue, magenta, cyan, gray and black
corresponding to $-1<\tau_{*}<1$,  $-2<\tau_{*}<2$, $-3<\tau_{*}<3$, $-4<\tau_{*}<4$, $-5<\tau_{*}<5$,  $-10<\tau_{*}<10$ and  $-15<\tau_{*}<15$, respectively.
The scale on the vertical axis is given in an arbitrary units. The absolute value can be determined once either $n_u$ or $n_{\gamma, u}$ are specified.
}
\label{Inug2e-1-2n5inte}
\end{figure*}

On the other hand, significant broadening is expected
when the rest mass energy is comparable or larger than that of the radiation at far upstream
($\xi_{u *} \lesssim 1$).
This corresponds to shocks that form around or above the saturation radius.
To gain some insight into how the radiation will be seen by an observer during the breakout of a RRMS under such conditions,
we plot, in Fig. \ref{Inug2e-1-2n5inte},  spectra  that were averaged over a certain 
physical interval $\Delta z$,
for models g2e-1n5 and g2e-2n5.
In addition to the angle integrated spectra that are computed by summing up the contribution of photons 
in all directions ($4 \pi$ steradians), we also display cases where
only the photons in a half hemisphere ($2 \pi$ steradians) propagating along ($\theta < \pi /2$) and
 against ($\theta > \pi / 2$) the flow are summed up.
The former (latter) represents the spectra emitted during the
breakout of the reverse (forward) shock that is advancing
relativistically in the central engine frame.
Hereafter we (loosely) refer to the cases that correspond to photons propagating along and against the flow as
 reverse and forward shock, respectively.  We emphasize that oblique shocks, that are likely to form near the photosphere,
 are also referred to here as reverse shocks, as their radiation escapes to infinity along the flow\footnote{Note that upon appropriate 
 Lorentz transformation oblique shocks can be transformed into perpendicular shocks.}.  We further point out that the spectra exhibited in Fig. \ref{Inug2e-1-2n5inte} 
 are computed in the
rest frame of the shock.
  In GRBs this frame moves at a high Lorentz factor with respect
 the observer, and so the observed emission is strongly beamed.   Thus, photons moving along the flow may significantly  contribute 
 to the observed spectrum also in forward shocks, depending on viewing  angle.  Our definition of "forward" and "reverse" in regards to the integrated spectrum
 is merely for explication.

As expected, there is a prominent hard component extending above the peak
 in the case of emission from a reverse shock.
It is produced by bulk Comptonization around the RRMS transition layer.
The spectrum emitted by a forward shock, on the other hand, lacks such 
a component (although it is broader than an exponential cutoff), since the high energy photons produced by  bulk Comptonization
move preferentially along the bulk flow.
In both cases, the portion of the spectrum below the  peak is softer (broader) than a thermal spectrum.
 This is  due to the moderately bulk Comptonized component in which energy gain by scattering is not
 so significant,
 as well as due to the superposition of thermal-like spectra emitted from the upstream and downstream regions.
 A substantial hardening is also seen 
 at the lowest energies (below the thermal peak of the radiation upstream), 
 since  none of the above mentioned broadening effects can play a role.

In comparison with  observations, the spectral slopes below and above the peak energy
 in reverse shocks
fall well within the range of detected values. 
For example, the reverse shock in model g2e-2n5 has low energy and  high energy photon indices ($d{\log}I_\nu/d{\log}\nu - 1$)
in the range  $-1.5 \lesssim \alpha \lesssim  -1$ and $-3 \lesssim \beta \lesssim  -2.5$
for the cases shown in  Fig. \ref{Inug2e-1-2n5inte} (middle right panel), which
are indeed in  good agreement with the observations \citep[e.g.,][]{YPG16}.
Similar values are found also for the low energy spectral index $\alpha$ in forward shocks.
On the other hand, unlike in reverse shocks, in forward shocks the spectrum above the peak shows a sudden drop off, and is
 incompatible with a power-law fit.
While this is in conflict with Band-like spectra,  
it is consistent with models that prefer an exponential-cutoff to fit observations, although it could be that those models are misled by an artifact of poor photon statistics at high energies \citep[e.g.,][]{KPB06,GBP12}.

It should be noted, however, that in general the shape of the  spectra emitted from a certain fluid shell
vary with the width of the shell, or in other words, with spatial interval $\Delta z$ over which they are averaged.
As we extend the length of this interval,  the contribution 
 from the far downstream and/or upstream regions increases and, therefore,
 the average spectrum asymptotes to a
thermal spectrum.\footnote{More accurately, it asymptotes to the superposition of
 two thermal components 
that have far upstream and downstream temperatures. The relative strength is determined by the ratio of spatially integrated intensity in each region.}
The reason is that  the bulk Comptonized component is confined to the vicinity of the shock transition layer by virtue 
of  efficient downscattering of high energy photons by the downstream plasma.
This means that  in case of a single shock that formed at a distance below the photosphere which is much 
larger than the width of the shock transition layer, 
while the signal around
 the time when the shock reaches photosphere
can  be highly non-thermal, the time integrated spectrum, that is, the spectrum integrated over the entire duration of burst,
would appear quasi-thermal. 
The broadening at sufficiently low energies might still prevail even then,
since its energy exchange rate with electrons is slower than that of the high energy photons.
Hence, the low energy broadening has a larger
 chance to be observed in the overall spectrum.

A more detailed analysis of the properties of photospheric GRB emission requires some knowledge about
how shocks are distributed within the outflow, which in turn determines the relative importance of
each emission region.
This is set by the nature of the central engine as well as by the environment into which the 
outflow is propagating.
Moreover, our calculations are restricted to infinite, steady shocks in planar geometry.
While our analysis can describe the shocks at regions well beneath the photosphere,
it cannot adequately address the breakout phase during which the photons diffuse out from the system.
One must take into account
 the drastic change in the shock structure during breakout  \citep{B17, GNL17} for a more accurate analysis of the released emission.
To that end, dynamical calculations must be performed which is beyond the scope of the present study.
Nevertheless, we emphasize that our steady-state simulations confirm that a
 broad, non-thermal spectrum is an inherent feature of RRMSs
 which  should also be present at the breakout phase.
Although more sophisticated computations are necessary for a firm conclusion, we suggest that
sub-photospheric shocks may provide a possible explanation for the non-thermal shape in the observed prompt emission spectra 
of GRBs.

\section{Summary and conclusions}
\label{conclusion}

We performed Monte-Carlo simulations of relativistic radiation mediated shocks for a broad range of upstream conditions.  Since photon generation is not
included in the current version of the code our results are applicable only to photon-rich shocks, for which the shock is supported by scattering of 
back streaming photons that were advected by the upstream flow.    To gain insight into the physical processes that shape the structure and spectrum of the shock,
 the results of the simulations are compared with analytic results whenever possible.    Our simulations confirm that the transition from photon rich to photon starved 
 regime occurs when the photon-to-baryon number ratio far upstream satisfies $\tilde{n}\simeq (m_p/m_e)\gamma_u$, as expected from an analytic comparison of the advection rate and the photon generation rate by the downstream plasma.   At this critical value the downstream temperature approaches
 the saturation value, roughly $200$ keV,  at which it is regulated by vigorous pair creation \citep{BKAW10}. 
 At sufficiently higher values of $\tilde{n}$ the 
 downstream temperature is much lower, pair loading is significantly reduced, and the shock is supported by the advected photons.

 We find that the deceleration of the bulk plasma occurs over a scale of
 a few pair loaded Thomson depths, with only a weak dependence on upstream 
 conditions; the actual physical scale may be much smaller 
 in cases where vigorous pair production ensues.  The shock width increases, but only slightly,
 when the relative contribution of high energy photons, that are scattered in the KN regime,
 becomes larger.   This is in difference to photon starved shocks in which the shock width is 
 essentially governed by KN effects \citep{BKAW10, GNL17}. 
%
  We also find that in the photon rich shocks we studied,
 the temperature of the plasma is determined almost
 solely by the Compton equilibrium throughout the shock, owing to
 the high photon-to-baryon number ratio ($\tilde{n}\gg 1$),
 with the exception of the immediate downstream temperature of the subshock whenever it is present.
Apart from these common features, our simulations indicate
 that the properties of the shock and its emission has a notable dependence on the 
 upstream parameters, $\xi_{u *}$, $\tilde{n}$, and $\gamma_u$.  Below we summarize our main findings:

\begin{itemize}
\item  

 When the energy density of the radiation far upstream largely exceeds the rest mass energy density ($\xi_{u *} \gg 1$), 
 the net increase in the radiation energy
 across the shock is small.   The dominance of the radiation renders the Lorentz factor profile smooth and broad;
 any attempt of steepening is readily smeared out by the large radiation drag acting upon the plasma.
Since the plasma cannot affect  much the radiation, the photon distribution is well described by a
 Wien distribution with a temperature that is equal to that of the local plasma temperature throughout the shock.
 In our fiducial model with $\tilde{n}=10^5$ the large value of $\xi_{u *}$ renders the temperature high enough 
 to induce significant pair production.
 The resulting population of pairs in this case can be
 well approximated by the Wien equilibrium.

\item  The situation changes drastically when $\xi_{u *} \lesssim 1$.
In this regime the radiation inside the shock becomes highly anisotropic,
and a significant fraction of the upstream bulk energy  is converted, via 
bulk Comptonization of counter streaming photons, to high-energy photons.
The consequent photon spectra exhibit a broad, non-thermal component 
that extends up to an energy of $\sim \gamma_u m_e c^2$.  
 The spectrum inside the shock becomes harder for lower values of  $\xi_{u *}$,
 leading to enhanced pair creation by virtue of the increased number of 
 photons with energies in excess of the pair production threshold.   
 The large pair enrichment in models with high $\gamma_u$ and low $\xi_{u*}$ 
 gives rise to a signature of the 511 keV annihilation line in the spectrum.

\item  It is also found that for sufficiently low  values of $\xi_{u *}$
 a weak subshock appears,
 although we were not able to identify its physical origin.
 Its effect on the  overall structure and emission of the shock is negligible,
 since it only dissipates a small amount of the total bulk kinetic energy.
 Thus, in practice the presence of the subshock is unimportant for the 
 overall analysis of RRMS, and it is merely of academic interest. 
 The above discussion excludes the cases with $\tilde{n}\simeq m_p/m_e$ (models g2e-1n3 and g2e-2n3), that 
delineate a transition between photon rich and photon starved RRMS, and for
which a strong subshock was found.  As explained above, the presence of a strong subshock in those models
is an artifact that stems from the omission of  photon production process in our code.  Once included, this subshock 
should become weak \citep{BKAW10}. 

 \item  In order that advected photons will be able to extract the entire upstream bulk energy, the number of advected photons
 per baryon  should exceed $m_p/m_e$ (i.e., $\tilde{n}>m_p/m_e$).  
 As stated above,  this value marks the transition between photon rich and photon starved shocks. 
 In the photon rich regime the value of $\tilde{n}$ merely determines the downstream temperature, that 
 scales as $T\propto \tilde{n}^{-1}$.  As a consequence, the pair production rate depends sensitively on the 
 value of $\tilde{n}$.

\item  The dependence on the bulk Lorentz factor $\gamma_u$ is relatively monotonic
 compared to the other two parameters.
 As the Lorentz factor increases, the
 maximum photon energy attainable through
 bulk Comptonization increases as $\sim \gamma_u m_e c^2$.
 Thus, the resulting photon spectra extends to higher energies.
 The  emergence of high energy photons that
 can diffuse back  to larger distances in the upstream region 
 also leads to an increase in the shock width,  in the peak
 temperature in the heating precursor, and in the density of pairs inside the shock.

\end{itemize}

We also considered the application to GRBs, and have
 shown that spectra  
 compatible with the observations can be produced within RRMSs.
 In particular, we demonstrate that
 the significant  spectral broadening 
 occurring in RRMS with $\xi_{u *} < 1$ can reproduce
 the typical Band-like spectrum.
 This result suggests that RRMS may be responsible, at least in part,
 for the non-thermal features found in the prompt emission spectra.
 However, our analysis is limited to infinite, planar shocks, and cannot account
 for the change in shock structure and emission caused by photon escape during
 the breakout phase, that may alter the observed spectrum.   We intend to carry out 
 detailed analysis of  breakout emission in a future work.

\section*{Acknowledgments}
We thank A. Beloborodov, I. Vurm, C. Lundman, D. Ellison and E. Nakar for fruitful discussions.
This work is supported by the Grant-in-Aid for
Young Scientists (B:16K21630) from The Ministry of Education, Culture, Sports, Science and Technology (MEXT).
Numerical computations and
data analysis were carried out on XC30 and PC cluster at
Center for Computational Astrophysics, National Astronomical
Observatory of Japan and at the Yukawa Institute Computer Facility. 
This work is supported in part by the Mitsubishi Foundation, a RIKEN pioneering project `Interdisciplinary Theoretical Science (iTHES)' and 'Interdisciplinary Theoretical \& Mathematical Science Program (iTHEMS)'.
AL acknowledges support by a grant from the Israel Science Foundation no. 1277/13

\appendix
\section{MONTE-CARLO RADIATION TRANSFER CALCULATION}
The Monte-Carlo  code used in this study handles transfer of photons in a medium at which Compton scattering, 
pair production, and pair annihilation takes place.
We iteratively perform many calculation runs, in order to find
the steady-state shock profile. 
In each iterative step, the photon transfer is solved under a given 
profile of number density, $n$ and $n_{\pm}$, temperature, $T$,
and velocity, $\beta$ of the plasma as follows:
%
We track the propagation of packets which are ensemble of photons that have identical 4-momentum, $P_{\gamma}^{\mu}=(h\nu/c, (h\nu/c) {\bf n})$, where ${\bf n}$ denotes the unit 3-vector along the propagation direction.
The photon packets are injected at the inner and outer boundaries which are located at the far upstream and downstream regions, respectively.
In addition, pair annihilation processes adds photons into the calculation domain.
After the injection, 
the evolution of the injected packets 
are computed until they  reach the  boundaries of the calculation domain or become absorbed via pair production process.
During the propagation, they are subject to multiple scatterings by the pair plasma.
Between each scattering event, the packet travels in the direction along the 3-vector ${\bf n}$.
After the scatterings, their
 4-momenta are updated based on the differential cross section of Compton scatterings, and the propagation direction
 is changed to newly determined ${\bf n}$.
Once we finish the calculation,
quantities such as total energy-momentum flux and distribution function of photons are evaluated by sampling all of the simulated packets at each grid point,

For each photon packet,  distances prior to the scattering and absorption events are determined by drawing the corresponding optical depths, $\delta \tau_{\rm sc}$ and $\delta \tau_{\pm}$.
The probability for the selected optical depth to be in the range of [$\delta \tau, \delta \tau + d\tau$] is given by ${\rm exp}(- \delta \tau)d\tau$.
From the given optical depths, path lengths to the scattering or absorption events in the laboratory  frame (shock rest frame) 
are determined from the  integration of opacity along the ray of photons, which can be expressed as
\begin{eqnarray}
\label{dtau}
\delta \tau = \int^{l}_0 {\cal D}^{-1}  \alpha^{'} dl, 
\end{eqnarray}
where ${\cal D} = [\gamma (1 - \beta  {\rm cos}\theta)]$ is the 
 Doppler factor. 
Here $\alpha^{'}$ denotes the opacity for the corresponding process in the comoving frame of the plasma which can be evaluated from the local physical conditions (see below for detail).

Below, we summarize how the injection
of photon packets as well as  scattering and absorption processes are treated in our code. 
Hereafter, we label quantities that are measured in the comoving frame
of the bulk plasma with the superscript prime symbol.

\subsection{Boundary Conditions}
\label{boundary}
At the boundaries located far upstream and downstream, we assume the photons are isotropic in the comoving frame and  have 
a Wien distribution characterized by the  local plasma temperature. 
Therefore,  the photon flux density
 at the boundary  in the laboratory frame is a function of the
 photon number density and temperature, and can be written as
\begin{eqnarray}
 \frac{dN_{\gamma}}{dt d\nu d\Omega dS} = {\cal D}^{2} \frac{dN_{\gamma}}{dt^{'} d\nu^{'} d\Omega^{'} dS^{'}} , 
\end{eqnarray}
where 
\begin{eqnarray}
 \frac{dN_{\gamma}}{dt^{'} d\nu^{'} d\Omega^{'} dS^{'}} = \frac{n_{\gamma}}{8 \pi}
                               \left(\frac{h}{k T} \right)^3 \nu^{'2} {\rm exp}\left( - \frac{h \nu^{'}}{k T}\right) . 
\end{eqnarray}
Thus, for a given range of solid angles  $d\Omega$ and frequencies $d\nu$, 
 $\frac{dN_{\gamma}}{dt d\nu d\Omega dS} (n_{\gamma, {pack}})^{-1} {\rm cos} \theta d\Omega d\nu$ gives the injection rate of  the packet number per unit area of the boundary surface, where $n_{\gamma, {\rm pack}}$ is the number of photons contained in a single packet.

\subsection{Pair annihilation}
\label{pairann}
The pair annihilation rate per unit volume
 is evaluated as a function of the pair number density and temperature:
\begin{eqnarray}
\left( \frac{dN_{\pm}}{dt dV} \right)_{\rm ann}
 = - (n + n_{\pm}/2) (n_{\pm}/2) c \sigma_T r_{\pm} (\Theta),
\label{Qann}
\end{eqnarray}
where
 $\sigma_T$ is the Thomson cross section.
Here 
 $r_{\pm}$ is an analytical function introduced by \citet{BKAW10} 
 based on the formula shown in \citet{S82}, which is given by
\begin{eqnarray}
r_{\pm} = \frac{3}{4} \left[1 + \frac{2 \Theta^2}{{\rm ln}(2 \eta_E \Theta + 1.3)} \right]^{-1},
\end{eqnarray}
where $\eta_E = e^{-\gamma_E} \approx 0.56146$ and $\gamma_E \approx 0.5772$ is the Euler's constant.
It is noted that  the above quantity  is Lorentz invariant (i.e., $\frac{dN}{dt dV} = \frac{dN}{dt^{'} dV^{'}}$).

As for the energy spectrum of the  photons produced via pair annihilation,
we use an fitting formula given in \citet{SLP96}, which approximates the exact emissivity in a wide range of temperatures.
By normalizing the given function to be consistent with the Equation (\ref{Qann}), it can be written as
\begin{eqnarray}
\label{VV}
\left( \frac{dN_{\gamma}}{dt d\nu d\Omega dV} \right)_{\rm ann}   =
 {\cal D} \left( \frac{dN_{\gamma}}{dt^{'} d\nu^{'} d\Omega^{'} dV^{'}} \right)_{\rm ann}
\end{eqnarray}
where 
\begin{eqnarray}
\label{VVcom}
 \left(  \frac{dN_{\gamma}}{dt^{'} d\nu^{'} d\Omega^{'} dV^{'}} \right)_{\rm ann} ~~ = ~~~~~~~~~~~~~~~~~~~~~~~~~~~~~~~~~~~~~~~~~~~~~~~~~~~~~~ & \\ 
  \left\{ \begin{array}{ll} 
         Q_1 \Theta^{0.5} x_{\nu}^{'3/2} {\rm exp}\left(- \frac{x_\nu^{'} + x_\nu^{'-1}}{\Theta} \right) \frac{C(x_\nu^{'} \Theta)}{K_2(1/\Theta)^2}  &~~
         {\rm for}~~  x_\nu^{'} \Theta \leq 20 ,  \\
        Q_2 x_\nu^{'} ( {\rm ln}4 \eta_E x_\nu^{'} \Theta - 1  ) {\rm exp}\left( -\frac{x_\nu^{'}}{\Theta} \right) \frac{C(x_\nu^{'} \Theta)}{K_2(1/\Theta)^2}  &~~
         {\rm for}~~ x_\nu^{'} \Theta > 20 . \\ 
 \end{array} \right .\nonumber
\end{eqnarray}
Here  $x_{\nu}^{'} = h \nu^{'} / (m_e c^2)$, 
and
 $K_2$ denote the  
 2nd order modified Bessel function of the second kind.
In evaluating the Bessel function, we used the approximate formula 
\begin{eqnarray}
K_2(1/\Theta)^2 & = & 4 \Theta^4 {\rm exp}\left(-\frac{2}{\Theta} \right) \\
                & \times &  [1 + 2.0049 \Theta^{-1} + 1.4774 \Theta^{-2}
                   + \pi (2\Theta)^{-3} ], \nonumber
\end{eqnarray}
which is also given in \citet{SLP96}.
The function $C$ is an analytical function given by 
\begin{eqnarray}
C(y)   ~~~ = ~~~~~~~~~~~~~~~~~~~~~~~~~~~~~~~~~~~~~~~~~~~~~~~~~~~~~~~~~~~~~~~~~~~~~~~~~~~~~~~~~~  \\
  \left\{ \begin{array}{ll}
       \frac{1 + 6.8515487 y + 1.4351694 y^2 + 0.001779014 y^3}{1 + 4.63115589 y + 1.5253007 y^2 + 0.04522338 y^3}&
         {\rm for}~  y \leq 20 ,  \\
      1 + 2.712 y^{-1} - 55.6 y^{-2} + 1039.8 y^{-3} - 7800 y^{-4}  &
         {\rm for}~ y > 20 . \\
 \end{array} \right. \nonumber
\end{eqnarray}
The normalization factors $Q_1$ and $Q_2$ are determined  
from the condition
 $\int \int \frac{dn_{\gamma}}{dt d\nu d\Omega dV} d\nu d\Omega = -  \left(\frac{N_{\pm}}{dt dV} \right)_{\rm ann}$.
In our code,   for a given range of solid angles  $d\Omega$ and frequencies $d\nu$, 
 $\int \int \left( \frac{dN_{\gamma}}{dt d\nu d\Omega dV} \right)_{\rm ann} (n_{\gamma, {pack}})^{-1}
 d\Omega d\nu$ gives the injection rate of  the packet number per unit volume in the calculation domain.

\subsection{Compton scatterings}
\label{Compton}
In evaluating the opacity of photons to Compton scattering, 
we fully take into account the thermal motion of the plasma and Klein-Nishina effects.
As a function of the photon frequency,  local density of pairs and temperature,
it is calculated as
\begin{eqnarray}
 \alpha_{\rm sc}^{'}(\nu^{'}) = \int \int \int F_{\rm sc}({\bf P_e}, T, \nu^{'})  d{\bf P_e}^3,
\end{eqnarray}
where
\begin{eqnarray}
 F_{\rm sc}({\bf P_e}, T, \nu^{'}) = (1 - \beta_e {\rm cos}{\theta_{e \gamma}}) (n + n_{\pm}) f_{B}({\bf P_e}, T) \sigma_{\rm sc}(\nu^{''}) ,
\end{eqnarray}
and
\begin{eqnarray}
 f_B(P_e, T) = \frac{1}{4\pi (m_e c)^3 \Theta  K_2(1/\Theta)}  {\rm exp}\left(- \frac{E_e}{kT} \right) ,
\end{eqnarray}
is  the Maxwell-J$\ddot{\rm u}$ttner distribution function and
\begin{eqnarray}
\sigma_{\rm sc}(\nu)  & = &   \frac{3}{4}\sigma_T 
 \biggl[ \frac{1 + x_{\nu}}{x_{\nu}^3}
       \left\{ \frac{2 x_{\nu} (1+x_{\nu})}{1+2 x_{\nu}}
          - {\rm ln}(1+2x_{\nu}) \right\}~~~~~~~~~~~~~~~ \\
 &  + &  \frac{1}{2x_{\nu}} {\rm ln}(1+2 x_{\nu}) - \frac{1 + 3x_{\nu}}{(1+2x_{\nu})^2}
  \biggr] ,  \nonumber
\end{eqnarray}
is the total cross section for Compton scattering.
Here $\nu^{''}$ denotes the frequency in the rest frame of pairs.
The quantities
 ${\bf P_e}$, $\beta$ and $\theta_{e \gamma}$ are, respectively,
the spatial components of the 4-momentum, the 3-velocity measured in units of the light speed, and
 the angle between the photon and pair directions measured in the comoving frame, and
 $P_e = |{\bf P_e}|$ and $E_e = ( (m_e c^2)^2 + (P_e c)^2 )^{0.5}$.

By plugging in the evaluated opacity in Equation (\ref{dtau}), we  determine the distance for the photon to propagate before scattering.
Once the position of the scattering event is determined, we choose the 
4-momentum of a thermal pair that will interact with the photon.
The probability for the pair within a range $d{\bf P_e}^3$
 to be drawn is given by $ F_{\rm sc} d{\bf P_e}^3 / \alpha_{\rm sc}$.
Then we transform the 4-momentum of photons to the rest frame of the 
chosen electron/positron and
 determine the 4-momentum  after the scattering based on
 the probability given by the
 differential cross section of Compton scattering:
\begin{eqnarray}
\frac{d\sigma_{\rm sc}}{d\Omega}=
 \frac{3}{16\pi} \frac{\nu_1^{2}}{\nu^2}
\left(
\frac{\nu}{\nu_1} + \frac{\nu_1}{\nu} - {\rm sin}^2 \theta_{\rm sc}
\right) ,
\end{eqnarray}
where $\nu_{1} = \nu [1 + x_{\nu}(1- {\rm cos}\theta_{\rm sc})]^{-1}$ is
the frequency after the scattering and $\theta_{\rm sc}$ is
the angle between the propagation directions of the incident and scattered photon.
Finally we transform back the 4-momentum of  the scattered photon
 to  the laboratory frame
and repeat the above cycle  until the packet is either absorbed or reaches the surface of the computation boundaries.


\subsection{Pair production}
For a given 4-momentum of incident photon in the comoving frame,
 $P_{\gamma}^{' \mu}=(h\nu^{'}/c, h\nu^{'}/c {\bf n^{'}})$,
the opacity for the pair production  is calculated as
\begin{eqnarray}
\alpha_{\gamma \gamma}^{'}(\nu^{'}, {\bf n^{'}}) = \int \int \int 
(1- {\rm cos}\theta_{\gamma \gamma})
 f_{\gamma}^{'}({\bf \tilde{P}}_{\gamma}^{'}) \sigma_{\gamma \gamma}(\nu^{'}, \tilde{\nu}^{'}, \theta_{\gamma \gamma}) {\bf \tilde{dP}_{\gamma}}^3,~~~
\end{eqnarray}
where ${\bf \tilde{P}_{\gamma}} = h \tilde{\nu}/c \tilde{{\bf n}}$ and
$\theta_{\gamma \gamma} $
denote the spacial component of the target photon 4-momentum and 
the angle between the propagation directions of the incident and target photons, respectively.
Here $ f_{\gamma}^{'}({\bf \tilde{P}}_{\gamma}^{'}) $ is the distribution function of photons in the comoving frame. 
The cross section for the interaction, taken from \citet{GS67}, is given by 
\begin{eqnarray}
&& \sigma_{\gamma \gamma}(\nu, \tilde{\nu}, \theta_{\gamma \gamma}) = 
\frac{3}{16}
  \sigma_T (1- \beta_{\rm cm}^2) ~~~~~~~~~~~~~~~~~~~~~~~~~~~~~  \\
   & \times &  \biggl[   (3 - \beta_{\rm cm}^4) 
  {\rm ln}\left( \frac{1 + \beta_{\rm cm}}{1 - \beta_{\rm cm}} \right) 
    - 2 \beta_{\rm cm} (2- \beta_{\rm cm}^2) \biggr] 
  ~~{\rm for}~~ \nu \geq \nu_{\rm thr}, ~~~~~~ \nonumber
\end{eqnarray}
and $\sigma_{\gamma \gamma} = 0$ for  $\nu < \nu_{\rm thr}$, where
 $h\nu_{\rm thr} = 2 m_e c^2 [h \nu (1 - {\rm cos}\theta_{\gamma \gamma}) ]^{-1}$ is the threshold energy for pair production and
$\beta_{\rm cm}= \sqrt{1 - 2 m_e^2 c^4/ [(1-{\rm cos}\theta_{\gamma \gamma})h^2\nu \tilde{\nu}]} $
 is the velocity of the pairs in the center of momentum frame.
As in the case of Compton scattering, the distance for the photon to propagate
 before being absorbed via the pair production process is computed by substituting the above opacity in Equation (\ref{dtau}).

In our code,
the local photon distribution function
 $ f_{\gamma} $ is determined by
  recording photon packet in each grid point during a single run of the simulation.
Since we cannot a-priori know the distribution of the current run before calculation,  in each run, we use the 
photon distribution obtained in the previous step in evaluating the opacity.
To ensure the self-consistently of our calculation,
we continue the iteration until the difference between the
 photon distribution evaluated in the current and previous run becomes sufficiently small.


\section{CALCULATION OF THE PAIR DENSITY PROFILE}
\label{steadypair}
In our code, independent from the energy-momentum conservation, we must
determine the  pair density profile which 
satisfy the steady state condition:
\begin{eqnarray}
\gamma \beta c \frac{d(\gamma n_{\pm}\beta)}{dz} = \left( \frac{dN_{\pm}}{dt dV} \right)_{\rm ann} +  \left( \frac{dN_{\pm}}{dt dV} \right)_{\rm pr},  
\end{eqnarray}
where $dz$ denotes the distance element and 
$( \frac{dN_{\pm}}{dt dV} )_{\rm pr}$ is the pair production rate.
While the annihilation rate is calculated based on the local quantities as shown in Equation (\ref{Qann}),
the production rate is evaluated by summing up the number of packets that 
are absorbed during the propagation in each grid points.
During the iteration, pair density profile is modified from the previous iteration step in order to
minimize the deviation from the above condition.
We continue the iteration until the error becomes sufficiently small.

\subsection{Wien equilibrium}
\label{WIENeq}
When the thermal temperature exceeds $k T \sim 30~{\rm KeV}$,
copious pairs are produced and can reach the equilibrium
state where pair production and annihilation is balanced (Wien equilibrium).
In this case, the number density of pairs can be
derived as
\begin{eqnarray}
n_{\pm} = n \left( \sqrt{1 +  \left(\frac{n_{\gamma}}{n}\right)^2
   \left( \frac{K_2(\Theta^{-1})}{\Theta^2} \right)^2 } - 1 \right) .
\end{eqnarray}
In the limit of $n_{\gamma} \gg n$, 
 it asymptotes to
\begin{eqnarray}
n_{\pm} \approx n_{\gamma} 
             \frac{K_2(\Theta^{-1})}{\Theta^2} .
\end{eqnarray}
If the temperature is 
 non-relativistic,  $\Theta \ll 1$, the above equation
 can be further approximated as
\begin{eqnarray}
n_{\pm} \approx \sqrt{\pi / 2}~ \Theta^{-3/2} {\rm exp} (- \Theta^{-1}) .
\label{nonrelaWIEN}
\end{eqnarray}

\section{CALCULATION OF THE TEMPERATURE PROFILE}
\label{Tcal}
In principle, the plasma temperature profile can be self-consistently obtained by
 determining the plasma profile for which
energy-momentum conservations is satisfied.
It is, however, numerically difficult to constrain the temperature profile accurately 
from the condition $F_{m} = {\rm const}, F_{e}= {\rm const}$ using our iterative method.
This stems from the fact that  since the contribution of the
thermal energy is always much smaller than that of the rest mass energy ($\rho_{pl} \gg e_{pl}$),
small numerical errors can lead to large errors in the temperature.
%
On the other hand, the radiation field responds non-linearly to changes in temperature, thereby rendering 
the calculations unstable and preventing  convergence to the steady state solution.

In order to overcome this numerical difficulty, we impose an additional constraint that can be derived from the energy-momentum conservation equations.
The equation solves the evolution of  internal energy density in comoving frame  which is given by
\begin{eqnarray}
\gamma \beta c \frac{de_{pl}}{dz} & = & \left( \frac{dE^{'}}{dt^{'}dV^{'}} \right)_{\rm sc} +
                    \left( \frac{dE^{'}}{dt^{'}dV^{'}} \right)_{\rm ann} + 
		    \left( \frac{dE^{'}}{dt^{'}dV^{'}} \right)_{\rm pr} ~~~ \\
                 & + & 
                    (e_{pl} + P_{pl})\gamma \beta c \frac{dn}{dz}, \nonumber
\end{eqnarray}
where
 $( \frac{dE^{'}}{dt^{'}dV^{'}} )_{\rm sc}$,  $( \frac{dE^{'}}{dt^{'}dV^{'}} )_{\rm ann}$, 
 and $( \frac{dE^{'}}{dt^{'}dV^{'}} )_{\rm pr}$ denote
 the net heating/cooling rate of the plasma by Compton scattering, pair annihilation, and production, respectively.  
The last term corresponds to the contribution of adiabatic heating (cooling)
due to compression (expansion).
While the adiabatic cooling term is evaluated from the local density gradient,
$( \frac{dE^{'}}{dt^{'}dV^{'}} )_{\rm sc}$, 
$( \frac{dE^{'}}{dt^{'}dV^{'}} )_{\rm ann}$, and 
 $( \frac{dE^{'}}{dt^{'}dV^{'}})_{\rm pr}$,  
are evaluated by summing up the contributions of scattering, annihilation and production of photon packets in  each grid point.
By using the above equation, we could successfully obtain an accurate temperature profile.

In our calculations, 
the temperature is almost solely determined by the condition of 
Compton equilibrium (net heating and cooling by Compton scattering is balanced, that is., 
$ \left( \frac{dE^{'}}{dt^{'}dV^{'}} \right)_{\rm sc} \sim 0$.
This is owning to the fact that the heat capacity of the plasma is extremely 
small compared to that of the radiation field, since the photon-to-lepton density ratio
 is large $n_{\gamma}/(n+n_{\pm})\gg 1$.
As a result, 
 any small deviation from the equilibrium will
 inevitably washed out by the numerous photons 
 within a length scale much smaller than
 the Thomson mean free path $\tau \ll 1$.
Notable deviation from the Compton equilibrium
temperature is only seen at the immediate downstream region of subshock 
where instantaneous viscous heating occurs.
Note that the downstream of the subshocks in models g2e-1n3 and  g2e-2n3
has a large extent ($\tau_{*} \gtrsim 1$) where equilibrium is not established, owing to the vigorous pair enrichment that
renders the heat capacity of the plasma comparable to that of the radiation ($n_{\gamma} \sim n_{\pm}$).



\label{lastpage}

\end{document}